\documentclass[aip,pop,a4paper,reprint,floatfix,groupedaddress]{revtex4-1}

\usepackage{epsfig}
\usepackage{graphicx}
\usepackage{amsmath}
\usepackage{amssymb}
\usepackage{amsfonts}

\begin{document}

\title{Energy transfer in binary collisions of two gyrating charged particles in a magnetic
field}
\author{H.~B.~Nersisyan}
\altaffiliation{Permanent address: Institute of Radiophysics and Electronics, 378410
Ashtarak, Armenia}
\email{hrachya@irphe.am}
\author{G.~Zwicknagel}
\email{guenter.zwicknagel@physik.uni-erlangen.de}
\affiliation{Institut f\"{u}r Theoretische Physik II, Erlangen-N\"{u}rnberg Universit\"{a}%
t, Staudtstrasse 7, D-91058 Erlangen, Germany}
\date{\today}

\begin{abstract}
Binary collisions of the gyrating charged particles in an external magnetic field
are considered within a classical second-order perturbation theory, i.e.,~up
to contributions which are quadratic in the binary interaction, starting
from the unperturbed helical motion of the particles. The calculations are
done with the help of a binary collisions treatment which is valid
for any strength of the magnetic field and  involves all harmonics of
the particles cyclotron motion. The energy transfer is explicitly calculated for a
regularized and screened potential which is both of finite range and
nonsingular at the origin. The validity of the perturbation treatment is
evaluated by comparing with classical trajectory Monte Carlo (CTMC) calculations
which also allow to investigate the strong collisions with large energy and
velocity transfer at low velocities. For large initial velocities on the
other hand, only small velocity transfers occur. There the nonperturbative
numerical CTMC results agree excellently with the predictions of the perturbative treatment.
\end{abstract}

\maketitle

\section{\label{sec:intr}Introduction}

In the presence of an external magnetic field $\mathbf{B}$ the problem of
two charged particles cannot be solved in a closed form as the relative
motion and the motion of the center of mass are coupled to each other.
Therefore no theory exists for a solution of this problem that is uniformly
valid for any strength of the magnetic field and the Coulomb force between
the particles. The energy loss of ion beams and the related processes in a
magnetized plasmas which are important in many areas of physics such as
transport, heating, magnetic confinement of thermonuclear plasmas and
astrophysics are examples of physical situations where this problem arises.
This topic was studied starting with the classic papers in Ref.~\onlinecite{ros60}
using kinetic equation approach, where the binary collisions of the
particles are masked due to the velocity average in the collision operator.
Recent applications are the cooling of heavy ion beams by electrons~\cite%
{sor83,pot90,mes94,men08} and the energy transfer for heavy-ion
inertial confinement fusion.\cite{pro08}

Calculations have been performed for binary collisions (BC) between magnetized
electrons~\cite{sia76,gli92} and for collisions between magnetized electrons and
ions.\cite{gzwi99,zwi00,zwi02,zwi06,toe02,ner03,ner07,ner09} In the presence of a
magnetic field only the total energy $E$ of the interacting particles is conserved
but not the relative and center of mass energies separately. In addition,
the rotational symmetry of the system is broken,
and as a consequence only the component of the angular momentum parallel to the
magnetic field is a constant of motion. The apparently simple problem of charged
particle interaction in a magnetic field is in fact a problem of considerable
complexity and the additional degree of freedom of the cyclotron orbital motion
produces a chaotic system with two degrees of freedom.\cite{gut90,sch00,bhu02,ner07}

In this paper we consider the BC between two gyrating charged particles treating the
interaction (Coulomb) as a perturbation to their helical motions. For electron-heavy
ion collisions this has been done previously in first-order in the ion charge $Z$ for
an ion at rest \cite{gel97} and up to $Z^{2}$ for an uniformly moving heavy ion.
\cite{toe02,ner03,ner07,ner09}
Here we focus on second-order perturbation theory and its comparison with classical
trajectory Monte Carlo (CTMC) simulations. The present work considerably extends the
earlier studies in Refs.~\onlinecite{ner07,ner03,ner09} where the second-order energy
transfer was calculated for an electron-heavy ion BC neglecting the gyration of the
ion or assuming identical gyrating particles (e.g. electrons). But in some cases
of an interaction of magnetized electrons with light ions (e.g. protons, antiprotons) at
a rather strong magnetic field (e.g. in traps) the cyclotron motion of the ion cannot be
neglected.

The general expressions for the energy transfers are briefly discussed in Sec.~\ref{sec:s1}.
In Sec.~\ref{sec:s2} we discuss the perturbative methodology used in this work. Then
we turn to the explicit calculation of the second-order energy transfer
for two arbitrary gyrating particle collision and without any restriction on the magnetic
field. In Sec.~\ref{sec:s4} the results of the perturbative BC model are compared with CTMC
simulations. The results are summarized and discussed in Sec.~\ref{sec:disc}. The small
velocity limits of the energy transfers are derived in the Appendix.

\section{\label{sec:s1}Binary collision formulation}

For our description of BC we start with the equations of motion for two
charged particles moving in a homogeneous magnetic field and the related conservation laws
in general. Next the quantities of interest, the velocity transfer, and the energy transfer
of particles during the binary collision will be introduced and discussed, before we turn to
the solution of the equations of motion in the subsequent section. For further details we
refer to Refs.~\onlinecite{ner07,ner03,ner09}.

We consider two point charges with masses $m_{1}$ and $m_{2}$ and charges $q_{1}e$ and $q_{2}e$,
respectively, moving in a homogeneous magnetic field $\mathbf{B}=B\mathbf{b}$ and interacting
with the potential $q_{1}q_{2}e\!\!\!/^{2}U(\mathbf{r})$ with $e\!\!\!/^{2}=e^{2}/4\pi \varepsilon _{0}$.
Here $\varepsilon _{0}$ is the permittivity of the vacuum and $\mathbf{r}=\mathbf{r}_{1}-\mathbf{r}_{2}$
is the relative coordinate of the colliding particles.

In plasma applications the bare Coulomb interaction $U_{\mathrm{C}}(\mathbf{r}) = 1/r$ is shielded by the surrounding
plasma particles and the interaction may
be modeled by $U_{\mathrm{D}}(\mathbf{r})=e^{-r/\lambda }/r$. Relative velocities which exceed the
thermal velocity lead to an asymmetric interaction potential which in general considerably complicates
the theoretical description. It has been shown, however, that such a dynamic, highly asymmetric
interaction potential can be replaced with an effective spherically symmetric velocity-dependent
interaction $U_{\mathrm{D}}(\mathbf{r})$ with a velocity-dependent
screening length $\lambda $.\cite{zwi99,zwi00,gzwi02} We adopt these findings for our present
considerations and assume a spherically symmetric interaction with a given fixed screening length
in both our analytic expressions and the CTMC simulations. For the envisaged applications of our
given results, such as cooling forces, stopping power, etc., which typically involves an average
over the velocity distribution, the screening length has then to be replaced by an appropriately
chosen velocity-dependent one.

The quantum uncertainty principle prevents particles (for $q_{1}q_{2}<0$) from falling into the
center of these potentials. In a classical picture this can be achieved by regularization of
$U(\mathbf{r})$ at the origin. Such a regularized potential (pseudopotential) has been derived
from quantum statistical considerations.\cite{kel63,deu77} In our forthcoming investigations we
take the functional form of this short distance correction and, including as well the screening
contribution, hence use the interaction $U_{\mathrm{R}}(\mathbf{r})=(1-e^{-r/\lambdabar})e^{-r/\lambda }/r$.
It should be emphasized, however, that the use of this regularized interaction, where $\lambdabar$
is usually related to the (thermal) de Broglie wavelength,\cite{kel63,deu77} here essentially
represents an alternative implementation of the standard (lower) cutoff procedure needed to handle
the hard collisions in a classical perturbative approach where $\lambdabar$ is taken as a given
constant or as a function of the classical collision diameter (see Sec.~\ref{sec:s4}).

Introducing the cyclotron frequencies of the particles $\omega _{c\nu }=|q_{\nu }|eB/m_{\nu }$
(with $\nu =1,2$) we start with the classical equations of motion
\begin{equation}
{\dot{\mathbf{v}}}_{\nu }(t)-\varsigma _{\nu }\omega _{c\nu }\left[ \mathbf{v%
}_{\nu }(t)\times \mathbf{b}\right] =\varrho _{\nu }\frac{%
q_{1}q_{2}e\!\!\!/^{2}}{m_{\nu }}\mathbf{F[r}(t)] ,
\label{eq:ind1}
\end{equation}%
where $\varsigma _{\nu }=|q_{\nu }|/q_{\nu }$, $\varrho _{1}=1$, $\varrho
_{2}=-1$. Here $q_{1}q_{2}e\!\!\!/^{2}\mathbf{F}[\mathbf{r}(t)]$ $(\mathbf{F}%
=-\partial U/\partial \mathbf{r})$ is the force exerted by particle 2 on
particle 1. In the presence of an external magnetic field, the Lagrangian
and the corresponding equations of motion cannot be separated into parts
describing the relative motion $[\mathbf{r}=\mathbf{r}_{1}-\mathbf{r}_{2},%
\mathbf{v}=\dot{\mathbf{r}}]$ and the motion of the cm $[\mathbf{R}=(m_{1}%
\mathbf{r}_{1}+m_{2}\mathbf{r}_{2})/(m_{1}+m_{2}),\mathbf{V}=\dot{\mathbf{R}}%
]$, in general.\cite{sia76,zwi02,toe02,ner03,ner09,ner07}
Introducing the reduced and the total masses $1/\mu =1/m_{1}+1/m_{2}$, $%
M=m_{1}+m_{2}$, respectively, and recalling that $\mathbf{v}_{\nu }(t)=%
\mathbf{V}(t)+\varrho _{\nu }(\mu /m_{\nu })\mathbf{v}(t)$ the equations of
the relative and the cm motion are
\begin{eqnarray}
&&{\dot{\mathbf{v}}}-\omega _{3}\left[ \mathbf{v}\times \mathbf{b}\right]
=\omega _{2}\left[ \mathbf{V}\times \mathbf{b}\right] +\frac{%
q_{1}q_{2}e\!\!\!/^{2}}{\mu }\mathbf{F[r}(t)],   \label{eq:a1}  \\
&&{\dot{\mathbf{V}}}-\omega _{1}\left[ \mathbf{V}\times \mathbf{b}\right]
=\frac{\mu }{M}\omega _{2}\left[ \mathbf{v}\times \mathbf{b}\right] .
\label{eq:a2}
\end{eqnarray}%
The frequencies $\omega _{1}$, $\omega _{2}$ and $\omega _{3}$ are expressed
in terms of the cyclotron frequencies of the particles, $\omega
_{1}=(m_{1}\varsigma _{1}\omega _{c1}+m_{2}\varsigma _{2}\omega _{c2})/M$, $%
\omega _{2}=\varsigma _{1}\omega _{c1}-\varsigma _{2}\omega _{c2}$, $\omega
_{3}=(m_{2}\varsigma _{1}\omega _{c1}+m_{1}\varsigma _{2}\omega _{c2})/M$.
Note that the quantities $|\omega _{1}|$ and $|\omega _{2}|$ play the role
of the cm and relative cyclotron frequencies. The coupled, nonlinear
differential equations~\eqref{eq:a1} and \eqref{eq:a2} (or Eq.~\eqref{eq:ind1})
completely describe the motion of the particles. For solving the
scattering problem, they have to be integrated numerically for a complete
set of the initial conditions.

From Eqs.~\eqref{eq:a1} and \eqref{eq:a2} follow the conservation of the
parallel component of the cm velocity ${\mathbf{V}}(t)\cdot \mathbf{b}%
=V_{0\parallel }$ and the total energy but since, in general, the relative
and center-of-mass motions are coupled the relative $E_{\mathrm{r}}$ and cm $%
E_{\mathrm{cm}}$ energies are not conserved separately. An exception is the
case with $\omega _{2}=0$ (or $q_{1}/m_{1}=q_{2}/m_{2}$) where the energies $%
E_{\mathrm{r}}$ and $E_{\mathrm{cm}}$ are conserved separately.

In the general case the rate $dE_{\nu }/dt$ at which the energy $E_{\nu}=m_{\nu }v_{\nu }^{2}/2$
of particle $\nu $ changes during the collision with the other particle can be obtained by
multiplying the equation of motion for particle $\nu $ by its velocity $\mathbf{v}_{\nu }(t)$.
The integration of this rate over the whole collision yields the energy transfer\cite{ner03,ner07,ner09}
\begin{equation}
\Delta E_{1}=-iq_{1}q_{2}e\!\!\!/^{2} \int d\mathbf{k}U(\mathbf{k})\int_{-\infty }^{\infty }
[\mathbf{k}\cdot \mathbf{V}(t)] e^{i\mathbf{k}\cdot \mathbf{r}(t)}dt
\label{eq:a10}
\end{equation}%
assuming that for $t\to \pm \infty $, $r(t)\to \infty $ and $U[\mathbf{r}(t)]\to 0$. According
to the conservation of total energy we have $\Delta E_{2}=-\Delta E_{1}$, as it can be directly
seen from Eqs.~\eqref{eq:ind1} and \eqref{eq:a10}. In Eq.~\eqref{eq:a10} the force $\mathbf{F}%
(\mathbf{r})$ has been written using a Fourier transformation in space.

\section{\label{sec:s2}Perturbative treatment}

We now seek an approximate solution of Eqs.~\eqref{eq:ind1}-\eqref{eq:a2} by assuming the interaction
force between the particles as a perturbation to their free helical motion. For the case of electron-ion
(without cyclotron motion of the ion) and electron-electron scattering the corresponding considerations
and derivations are discussed in Refs.~\onlinecite{ner07,ner09}. We therefore focus here on the general
case.

We look for a solution of Eq.~\eqref{eq:ind1} for the variables $\mathbf{r}_{\nu },\mathbf{v}_{\nu }=%
\dot{\mathbf{r}}_{\nu }$, or  alternatively, of Eqs.~\eqref{eq:a1} and \eqref{eq:a2}) for the variables
$\mathbf{r},\mathbf{v}=\dot{\mathbf{r}}$ and $\mathbf{R},\mathbf{V}=\dot{\mathbf{R}}$, in a perturbative
manner \cite{ner07,ner09}, i.e.~$\mathbf{r}_{\nu }=\mathbf{r}_{\nu }^{(0)}+\mathbf{r}_{\nu }^{(1)}+...$
and $\mathbf{R}=\mathbf{R}_{0}+\mathbf{R}_{1}+... $, $\mathbf{r}=\mathbf{r}_{0}+\mathbf{r}^{(1)}+...$
Starting point is the zero--order unperturbed helical motion of two particles in the laboratory frame
with $\mathbf{v}_{\nu }^{(0)}(t)=\dot{\mathbf{r}}_{\nu }^{(0)}(t)$ and
\begin{eqnarray}
&&\mathbf{r}_{\nu }^{(0)}\left( t\right) =\widetilde{\mathbf{R}}_{\nu }+%
\mathbf{b}v_{0\nu \parallel }t   \label{eq:a27}   \\
&&+a_{\nu }\left[ \mathbf{u}_{\nu }\sin \left(
\omega _{c\nu }t\right) +\varsigma _{\nu }\left[ \mathbf{b}\times \mathbf{u}%
_{\nu }\right] \cos \left( \omega _{c\nu }t\right) \right] ,   \nonumber
\end{eqnarray}%
where $\mathbf{u}_{\nu}=(\cos\varphi_{\nu },\sin\varphi_{\nu})$ ($\varphi_{\nu }$ is the initial phase of
the particle $\nu $) is the unit vector perpendicular to the magnetic field, $v_{0\nu \parallel }$ and
$v_{0\nu\bot}\mathbf{u}_{\nu}$ (with $v_{0\nu \bot}\geqslant 0$) are the unperturbed velocity components
parallel and perpendicular to $\mathbf{b}$, respectively. Here $a_{\nu }=v_{0\nu \perp }/\omega _{c\nu }$
is the cyclotron radius of the particle $\nu $. It should be noted that in Eq.~\eqref{eq:a27} the variables
$\mathbf{u}_{\nu }$ and $\widetilde{\mathbf{R}}_{\nu}$ are independent and are defined by the initial conditions.
The unperturbed cm ($\mathbf{R}_{0}(t),\mathbf{V}_{0}(t)=\dot{\mathbf{R}}_{0}(t)$) and relative
($\mathbf{r}_{0}(t),\mathbf{v}_{0}(t)=\dot{\mathbf{r}}_{0}(t)$) coordinates and velocities can be easily
found from Eq.~\eqref{eq:a27}. In general, the cm and relative coordinates and velocities involve two harmonic
oscillations with different frequencies, amplitudes and phases. Therefore in the plane perpendicular to
$\mathbf{b}$ these quantities cannot be represented in the form of a simple cyclotron motion with constant
amplitudes and phases as in the case of electron-electron (or electron-positron) collisions.\cite{ner09}

The equation for the first-order velocity correction is given by
\begin{equation}
{\dot{\mathbf{v}}}_{\nu }^{(1)}(t)-\varsigma _{\nu }\omega _{c\nu }[\mathbf{v%
}_{\nu }^{(1)}(t)\times \mathbf{b}]=\varrho _{\nu }\frac{q_{1}q_{2}e\!\!%
\!/^{2}}{m_{\nu }}\mathbf{F[r}_{0}(t)] ,
\label{eq:a28}
\end{equation}%
where the solutions can be given by an integral involving the force $\mathbf{F[r}_{0}(t)]$ with $\mathbf{r}_{0}(t)%
=\mathbf{r}_{1}^{(0)}(t)-\mathbf{r}_{2}^{(0)}(t)$ and the unperturbed trajectory \eqref{eq:a27} by similar
expressions as Eqs.~(43)-(46) of Ref.~\onlinecite{ner03} (see also Ref.~\onlinecite{ner07,ner09}).

The first- ($\Delta E^{(1)}_{1}$) and the second-order ($\Delta E^{(2)}_{1}$) energy transfers of the particle 1
can be evaluated using general Eq.~\eqref{eq:a10}. Here we consider only the energy change $\Delta E^{(2)}_{1}$
since the angular averaged $\Delta E^{(1)}_{1}$ vanishes due to the symmetry reasons.\cite{ner03,ner07,ner09} We obtain
\begin{eqnarray}
&&\Delta E_{1}^{(2)}=q_{1}q_{2}e\!\!\!/^{2}\int d\mathbf{k}U(\mathbf{k}%
)\int_{-\infty }^{\infty } e^{i\mathbf{k}\cdot \mathbf{r}_{0}(t)}dt  \nonumber  \\
&&\times \{ [\mathbf{k}\cdot \mathbf{r}^{(1)}(t)][\mathbf{k}\cdot \mathbf{V}_{0}(t)]
-i[\mathbf{k}\cdot \mathbf{V}_{1}(t)] \} .  \label{eq:a37}
\end{eqnarray}%
Let us recall that here $\mathbf{r}^{(1)}(t)$\ and $\mathbf{V}_{1}(t)$ are the first-order relative coordinate and
the cm velocity corrections, respectively.

We now introduce the variable $\mathbf{s}=\mathbf{R}_{r\bot }$ which is the
component of $\mathbf{R}_{r}=\widetilde{\mathbf{R}}_{1}-\widetilde{\mathbf{R}%
}_{2}$ perpendicular to the relative velocity of the guiding centers of two
particles $v_{r\parallel }\mathbf{b}$, where $v_{r\parallel }=v_{01\parallel
}-v_{02\parallel }$. From Eq.~\eqref{eq:a27} we can see that $\mathbf{s}$ is
the distance of closest approach for the guiding centers of the two
particles' helical motion. For practical applications the energy change is
given by the average of $\Delta E_{1}$ with respect to the initial
phases of the particles $\varphi _{1}$ and $\varphi _{2}$ and the azimuthal
angle $\vartheta _{\mathbf{s}}$ of $\mathbf{s}$. This averaged quantity
is denoted by $\langle \Delta E_{1}\rangle $ in the forthcoming considerations.

To evaluate the second-order energy transfer $\Delta E_{1}^{(2)}$ we have to insert Eq.~\eqref{eq:a27}
and the solution of Eq.~\eqref{eq:a28} into Eq.~\eqref{eq:a37}. This quantity is then averaged with
respect to the initial phases of the particles $\varphi _{1}$ and $\varphi _{2}$ and the azimuthal angle
$\vartheta _{\mathbf{s}}$ of the impact parameter $\mathbf{s}$. The obtained angular integrals are evaluated
using the Fourier series of the exponential function. After averaging the energy transfer $\Delta E_{1}^{(2)}$
with respect to $\varphi _{1}$ and $\varphi _{2}$ the remaining part will depend on $\delta (k_{\parallel
}+k_{\parallel }^{\prime })$, i.e. the component of $\mathbf{k}+\mathbf{k}^{\prime }$ along the magnetic field
$\mathbf{b}$. This $\delta $-function enforces $\mathbf{k}+\mathbf{k}^{\prime }$ to lie in the plane transverse
to $\mathbf{b}$ so that $e^{i(\mathbf{k+k}^{\prime })\cdot\mathbf{R}_{r}}\delta (k_{\parallel }+k_{\parallel }%
^{\prime })=e^{i\mathbf{Q}\cdot \mathbf{s}}\delta (k_{\parallel }+k_{\parallel }^{\prime })$, where
$\mathbf{Q}=\mathbf{k}_{\bot }+\mathbf{k}_{\bot }^{\prime }$. The result of the angular averaging finally reads
\begin{eqnarray}
&&\langle\Delta E_{1}^{(2)}\rangle =\frac{\pi iq_{1}^{2}q_{2}^{2}e\!\!\!/^{4}%
}{\mu\left\vert v_{r\parallel}\right\vert }\int d\mathbf{k}d\mathbf{k}%
^{\prime}U\left(  \mathbf{k}\right)  U\left(  \mathbf{k}^{\prime}\right)
J_{0}\left(  Qs\right)  \nonumber\\
&& \times\delta(k_{\parallel}^{\prime}+k_{\parallel})\sum_{n,m=-\infty}%
^{\infty}\left(  -1\right)  ^{m+n}e^{i(\theta-\theta^{\prime})(n\varsigma
_{1}+m\varsigma_{2})}   \nonumber \\
&& \times\left(  k_{\parallel}V_{0\parallel}+\frac{\mu}{m_{2}}n\omega
_{c1}-\frac{\mu}{m_{1}}m\omega_{c2}\right)  \Bigg\{ \frac{2k_{\parallel}%
^{2}H_{n,m}(k_{\perp},k_{\perp}^{\prime})}{\left(  \zeta_{n,m}-i0\right)
^{2}}   \nonumber \\
&& -\frac{k_{\perp}^{\prime}k_{\perp}}{\zeta_{n,m}-i0}\bigg[  2iA H_{n,m}(k_{\perp},k_{\perp}^{\prime})\sin(\theta
-\theta^{\prime})  \nonumber\\
&&   +\frac{\mu}{m_{1}\omega_{c1}}\mathcal{H}_{n,m}(k_{\perp
},k_{\perp}^{\prime})+\frac{\mu}{m_{2}\omega_{c2}}\mathcal{P}_{n,m}(k_{\perp
},k_{\perp}^{\prime})\bigg]  \Bigg\} .  \label{eq:a42}
\end{eqnarray}
Here $H_{n,m}=J_{n}(k_{\perp}a_{1}) J_{n}(k_{\perp }^{\prime }a_{1})J_{m}(k_{\perp }a_{2})
J_{m}(k_{\perp }^{\prime }a_{2})$ and $\mathcal{H}_{n,m}=H_{n-1,m}-H_{n+1,m}$,
$\mathcal{P}_{n,m}=H_{n,m-1}-H_{n,m+1}$, $A=\mu\varsigma_{1}/m_{1}\omega_{c1}
+\mu\varsigma_{2}/m_{2}\omega_{c2}$, where $J_{n}$ are the Bessel functions of the $n$th order. Also $\zeta _{n,m}=n\omega
_{c1}+m\omega _{c2}+k_{\parallel }v_{r\parallel }$, $k_{\parallel }=\mathbf{k%
}\cdot \mathbf{b}$ and $\mathbf{k}_{\bot }$ are the components of $\mathbf{k}$
parallel and transverse to $\mathbf{b}$, respectively, $\tan \theta =k_{y}/k_{x}$.
The series representation \eqref{eq:a42} of the second-order energy transfer is
valid for any strength of the magnetic field.

For most applications it is also useful to integrate the averaged energy transfer, $\langle\Delta E_{1}^{(2)}\rangle $,
with respect to the impact parameters $s$ in the full two-dimensional (2D) space. We thus introduce an energy loss
cross section (ELCS) \cite{gzwi99,zwi00,ner03,ner07,ner09} through the relation
\begin{equation}
\sigma =\int_{0}^{\infty }\langle \Delta E_{1}^{(2)}\rangle sds.
\label{eq:a43}
\end{equation}%
As $\sigma $ results from the $s$-integration of the energy transfer \eqref{eq:a42} one obtains an expression
for $\sigma $ which represents an infinite sum over Bessel functions. Moreover, assuming regularized interaction
and performing $\mathbf{k}$ integration in $\sigma $ yields an infinite sum over modified Bessel functions (see,
e.g., an example for ion-electron collision in Ref.~\onlinecite{ner07}). For arbitrary axially symmetric interaction
potential similar expression is derived in the Appendix [see Eq.~\eqref{eq:apb1}]. However, for practical
applications it is much more convenient to use an equivalent integral representation of the ELCS which does not
involve any special function. This expression can be derived from the Bessel-function representation of $\sigma $
using the integral representation of the Dirac $\delta $ function as well as the summation formula for
$\sum_{n}e^{in\varphi}J_{n}^{2}(a)$.\cite{gra80} The energy transfer $\sigma $ after lengthy but straightforward
calculations then reads
\begin{equation}
\sigma =\int_{0}^{2\pi }\frac{d\varphi }{2\pi }\overline{\sigma }\left(
\varphi \right) ,\quad \overline{\sigma }(\varphi )=\overline{\sigma }%
_{\parallel }(\varphi )+\overline{\sigma }_{\bot }(\varphi ),
\label{eq:cross}
\end{equation}%
with
\begin{eqnarray}
&&\overline{\sigma }_{\parallel }\left( \varphi \right) =-\frac{\left( 2\pi
\right) ^{2}q_{1}^{2}q_{2}^{2}e\!\!\!/^{4}\lambda ^{2}V_{0\parallel }}{\mu
v_{r\parallel }^{3}}\int_{0}^{\infty }tdt\int d\mathbf{k}\left\vert U\left(
\mathbf{k}\right) \right\vert ^{2}  \nonumber  \\
&&\times \lbrack k_{\parallel }^{2}+k_{\perp }^{2}\phi \left( t\right)
]k_{\parallel }\sin \left( k_{\parallel }\lambda t\right) J_{0}\left[
2k_{\perp }\mathcal{R}\left( t\right) \right] ,  \label{eq:a43b}
\end{eqnarray}%
\begin{eqnarray}
&&\overline{\sigma }_{\perp }\left( \varphi \right) =-\frac{\left( 2\pi
\right) ^{2}q_{1}^{2}q_{2}^{2}e\!\!\!/^{4}\lambda }{m_{1}m_{2}v_{r\parallel
}^{4}}\int_{0}^{\infty }t^{2}dt\int d\mathbf{k}\left\vert U\left( \mathbf{k}%
\right) \right\vert ^{2}k_{\perp }^{2}  \nonumber \\
&&\times \cos \left( k_{\parallel }\lambda t\right) \bigg\{ \lambda ^{2}G_{1}\left( t\right) [k_{\parallel
}^{2}+k_{\perp }^{2}\phi \left( t\right) ]\frac{J_{1}\left[ 2k_{\perp }%
\mathcal{R}\left( t\right) \right] }{2k_{\perp }\mathcal{R}\left( t\right) }  \nonumber  \\
&& +G_{2}\left( t\right) J_{0}\left[ 2k_{\perp }\mathcal{R}\left( t\right) %
\right] \bigg\} ,  \label{eq:a43bb}
\end{eqnarray}%
where $\delta _{\nu}=|v_{r\parallel }|/\omega _{c\nu}$ are the relative pitches of the
particles helices, divided by $2\pi $ and $\eta _{\nu}=\lambda /\delta
_{\nu}=\omega _{c\nu}\lambda /|v_{r\parallel }|$. In Eqs.~\eqref{eq:a43b} and
\eqref{eq:a43bb} the time $t$\ is scaled in units $\lambda /|v_{r\parallel }|$,
where the length $\lambda $ is specified in the next section. In addition
in Eq.~\eqref{eq:cross} the energy transfer $\overline{\sigma }(\varphi )$
has been split into two parts which correspond to the cm motion along [$%
\overline{\sigma }_{\parallel }(\varphi )$] and transverse [$\overline{%
\sigma }_{\perp }(\varphi )$] to the magnetic field, where $\varphi =\varphi
_{1}-\varphi _{2}$ is the difference of the initial phases of the particles.
Also
\begin{eqnarray}
&&\phi \left( t\right) =\frac{\mu }{m_{1}}\frac{\sin \left( \eta _{1}t\right)
}{\eta _{1}t}+\frac{\mu }{m_{2}}\frac{\sin \left( \eta _{2}t\right) }{\eta _{2}t},  \label{eq:phi} \\
&&\mathcal{R}^{2}\left( t\right) =a_{1}^{2}\sin ^{2}\frac{\eta _{1}t}{2}%
+a_{2}^{2}\sin ^{2}\frac{\eta _{2}t}{2}  \nonumber \\
&&-2a_{1}a_{2}\sin \frac{\eta _{1}t}{2}\sin \frac{\eta _{2}t}{2}\cos \varphi ,  \label{eq:R}
\end{eqnarray}
\begin{eqnarray}
&&G_{1}\left( t\right) =m_{1}v_{01\perp }^{2}\frac{\sin \left( \eta
_{1}t\right) }{\eta _{1}t}-m_{2}v_{02\perp }^{2}\frac{\sin \left( \eta
_{2}t\right) }{\eta _{2}t}+2v_{01\perp }v_{02\perp }  \nonumber  \\
&&\times \cos \varphi \left( m_{2}\frac{\sin \frac{\eta
_{1}t}{2}}{\eta _{1}t}\cos \frac{\eta _{2}t}{2}-m_{1}\frac{\sin \frac{\eta
_{2}t}{2}}{\eta _{2}t}\cos \frac{\eta _{1}t}{2}\right) ,  \label{eq:G1} \\
&&G_{2}\left( t\right) =\mu v_{r\parallel }^{2}\frac{\cos \left( \eta
_{2}t\right) -\cos \left( \eta _{1}t\right) }{t^{2}}.  \label{eq:G2}
\end{eqnarray}%
Here the quantity $\mathcal{R}(t)$ represents an effective relative cyclotron radius of the two particles.
The second term in Eq.~\eqref{eq:a43bb} proportional to the function $G_{2}(t)$ is the contribution of the
cm velocity perturbation, see the second term of Eq.~\eqref{eq:a37}. This perturbation and hence this
corresponding term was absent in our previous considerations for electron-electron and electron-ion collisions.

An expression similar to Eq.~\eqref{eq:a42} (or Eqs.~\eqref{eq:cross}-\eqref{eq:a43bb}) has been obtained
for electron-electron collision\cite{ner09} and for electron-heavy ion (without cyclotron motion of the ion)
collision.\cite{ner03,ner07,ner09} In contrast to these specific cases Eqs.~\eqref{eq:a42} and
\eqref{eq:cross}-\eqref{eq:a43bb} are valid for the collision of two arbitrary charged particles. The results
obtained in Refs.~\onlinecite{ner03,ner07,ner09} can be easily derived from these general expressions. The
electron-electron case is recovered assuming that $m_{1}=m_{2}=m$, $q_{1}=q_{2}=q$, i.e., $\omega _{c1}=%
\omega _{c2}=\omega _{c}$, $\eta _{1}=\eta _{2}=\eta $, $\mu =m/2$. Then the term proportional to the function
$G_{2}(t)$ in Eq.~\eqref{eq:a43bb} vanishes and the remaining expressions coincide with the results of
Ref.~\onlinecite{ner09}. It should be noted that for identical particles (e.g. electrons) the effective
cyclotron radius is given by $\mathcal{R}(t) =a\sin (\eta t/2)$, where $a^{2}=a_{1}^{2}+a_{2}^{2}-2a_{1}a_{2}%
\cos \varphi $ is the cyclotron radius of the particles in the relative frame.\cite{ner09} For electron-heavy
ion collision we assume that the ion mass tends to infinity, $m_{2}\to \infty $, and $v_{02\perp }=0$ (ion
moves along the magnetic field direction). Therefore $\mu \to m_{1}=m$, $\omega _{c2}\to 0$ ($\eta _{2}\to 0$).
In this case the transverse ELCS vanishes and the longitudinal cross section coincides with the results of
Ref.~\onlinecite{ner09}.

Next we also consider the ELCS $\overline{\sigma }_{\parallel }(\varphi )$ and $\overline{\sigma }_{\bot }%
(\varphi )$ for vanishing cyclotron radii, $a_{1},a_{2}\to 0$, i.e. when initially the particles move along
the magnetic field ($v_{01\bot }=v_{02\bot }=0$). In this limit $\overline{\sigma }_{\bot }(\varphi )\neq 0$
as in the cases of two identical particles collision or ion-electron collision. For any axially symmetric
interaction potential, $U(\mathbf{k})=U(|k_{\parallel }|,k_{\perp})$, the straightforward integration in
Eqs.~\eqref{eq:a43b} and \eqref{eq:a43bb} yields
\begin{eqnarray}
&&\overline{\sigma }_{\parallel }\left( \varphi \right) =-\frac{%
2q_{1}^{2}q_{2}^{2}e\!\!\!/^{4}V_{0\parallel }}{v_{r\parallel }^{3}}\left[
\frac{\mathcal{F}_{0}\left( \delta _{1}^{-1}\right)}{m_{1}}
+\frac{\mathcal{F}_{0}\left( \delta _{2}^{-1}\right)}{m_{2}} \right] ,  \label{eq:van1} \\
&&\overline{\sigma }_{\perp }\left( \varphi \right) =\frac{%
2q_{1}^{2}q_{2}^{2}e\!\!\!/^{4}}{Mv_{r\parallel }^{2}}\left[ \mathcal{F}%
_{0}\left( \delta _{1}^{-1}\right) -\mathcal{F}_{0}\left( \delta
_{2}^{-1}\right) \right] ,  \label{eq:van2}
\end{eqnarray}%
where
\begin{equation}
\mathcal{F}_{0}\left( \kappa \right) =\frac{\left( 2\pi \right) ^{4}}{4}%
\int_{0}^{\infty }U^{2}\left( \kappa ,k_{\perp }\right) k_{\perp
}^{3}dk_{\perp } .
\label{eq:fun}
\end{equation}

The averaged energy transfer, Eq.~\eqref{eq:a42}, can be evaluated without further approximation
for any axially symmetric interaction potential. In this case the energy transfer can be represented
as the sum of all cyclotron harmonics as it has been done for ion-electron interaction in
Ref.~\onlinecite{ner07} and for electron-electron interaction in Ref.~\onlinecite{ner09}.

In the following we consider the regularized screened potential introduced in Sec.~\ref{sec:s1} with
\begin{equation}
U_{\mathrm{R}}(k_{\parallel },k_{\perp })=\frac{2}{(2\pi)^{2}}
\left( \frac{1}{k_{\perp }^{2}+\kappa ^{2}}-\frac{1}{k_{\perp
}^{2}+\chi ^{2}}\right) ,
\label{eq:a48}
\end{equation}%
where $\kappa ^{2}=k_{\parallel }^{2}+\lambda ^{-2}$, $\chi ^{2}=k_{\parallel }^{2}+d^{-2}$,
$d^{-1}=\lambda ^{-1}+\lambdabar ^{-1}$.

As we discussed above the energy transfer \eqref{eq:a42} must be integrated with respect
to the impact parameters $s$ for practical applications. For general interaction potential
this is given by Eqs.~\eqref{eq:a43}-\eqref{eq:a43bb}. In general for a study of the
convergence of the $s$-integrated energy transfers we note that the case with some value
$s=s_{c}$ is most critical for the convergence of the ELCS. This is intuitively clear as
the gyrating particles at $\vert a_{1}-a_{2}\vert <s<a_{1}+a_{2}$ may hit each other on such
a trajectory (see Ref.~\onlinecite{ner07} for some explicit examples). This should not matter
for the potential \eqref{eq:a48}, which has been regularized near the origin for exactly that
purpose.

For the present case of the regularized interaction potential, substituting Eq.~\eqref{eq:a48}
into Eqs.~\eqref{eq:a43b} and \eqref{eq:a43bb}, we obtain
\begin{eqnarray}
&&\overline{\sigma }_{\parallel }(\varphi ) =-\frac{q_{1}^{2}q_{2}^{2}e\!\!\!/^{4}
V_{0\parallel }}{\mu v_{r\parallel }^{3}} \int_{0}^{\infty }\frac{t^{2}dt}{R^{3}(t)}
\bigg\{ e^{-R(t)}\bigg[ \mathcal{F}_{1}[R(t),t,t]  \nonumber \\
&&+\frac{4}{\varkappa ^{2}-1}\mathcal{F}_{2}[R(t),t,t] \bigg]
+e^{-\varkappa R(t)}\bigg[ \mathcal{F}_{1}[\varkappa R(t),\varkappa t,t]  \nonumber \\
&&-\frac{4\varkappa ^{2}}{\varkappa ^{2}-1}\mathcal{F}_{2}[\varkappa R(t) ,\varkappa
t,t] \bigg] \bigg\} ,  \label{eq:a63}
\end{eqnarray}
\begin{eqnarray}
&&\overline{\sigma }_{\perp }(\varphi ) =-\frac{q_{1}^{2}q_{2}^{2}e\!\!\!/^{4}}%
{m_{1}m_{2}v_{r\parallel }^{4}}\int_{0}^{\infty }\frac{t^{2}dt}{R^{3}(t)}\Bigg\{ G_{1}(t)
\bigg\{ e^{-R(t)}   \nonumber \\
&& \times \left[ \mathcal{F}_{3}[R(t),t,t]+\frac{4}{\varkappa ^{2}-1}\mathcal{F}_{4}[
R(t),t,t]\right]+e^{-\varkappa R(t)}   \nonumber \\
&& \times \left[ \mathcal{F}_{3}[\varkappa R(t),\varkappa t,t]-\frac{4\varkappa ^{2}}{
\varkappa ^{2}-1}\mathcal{F}_{4}[\varkappa R(t),\varkappa t,t] \right] \bigg\}  \nonumber  \\
&&+G_{2}(t)\bigg\{ e^{-R(t)}\left[ \mathcal{F}_{5}[R(t),t] +\frac{4}{\varkappa ^{2}-1}\mathcal{F}_{6}%
[R(t),t]\right]  \label{eq:a63x} \\
&& +\frac{e^{-\varkappa R(t)}}{\varkappa ^{2}} \left[ \mathcal{F}_{5}[\varkappa R(t),\varkappa t] -%
\frac{4\varkappa ^{2}}{\varkappa ^{2}-1}\mathcal{F}_{6}[\varkappa R(t),\varkappa t] \right] \bigg\} \Bigg\} .  \nonumber
\end{eqnarray}
Here $R^{2}(t)=t^{2}+(4/\lambda ^{2})\mathcal{R}^{2}(t)$, $\varkappa
=\lambda /d=1+\lambda /\lambdabar $, and
\begin{eqnarray}
&&\mathcal{F}_{1}(R,\zeta ,t) =2+2R-R^{2}+[1-\phi (t)] \bigg[ R^{2}+R+1  \nonumber \\
&& -\frac{\zeta ^{2}}{R^{2}}(R^{2}+3R+3) \bigg] ,  \label{eq:a64} \\
&&\mathcal{F}_{2}(R,\zeta ,t) =R+1-\frac{1}{R^{2}}[1-\phi (t)] \bigg[ R^{3}+4R^{2}+9R+9  \nonumber \\
&& -\frac{\zeta ^{2}}{R^{2}}(R^{3}+6R^{2}+15R+15) \bigg] , \label{eq:a65} \\
&&\mathcal{F}_{3}(R,\zeta ,t) =2+2R-R^{2}+[1-\phi (t)] \bigg[ R^{2}-R-1 \nonumber \\
&&-\frac{\zeta ^{2}}{R^{2}}(R^{2}+3R+3) \bigg] , \label{eq:a66} \\
&&\mathcal{F}_{4}(R,\zeta ,t) =R+1-\frac{1}{R^{2}}[1-\phi (t)] \bigg[ R^{3}+2R^{2}+3R+3 \nonumber \\
&&-\frac{\zeta ^{2}}{R^{2}}( R^{3}+6R^{2}+15R+15) \bigg] , \label{eq:a67} \\
&&\mathcal{F}_{5}(R,\zeta ) =R^{2}\left[ 1-R+\frac{\zeta ^{2}}{%
R^{2}}(R+1) \right] , \label{eq:a66b} \\
&&\mathcal{F}_{6}(R,\zeta )=R^{2}+R+1-\frac{\zeta ^{2}}{R^{2}}(R^{2}+3R+3) .
\label{eq:a67b}
\end{eqnarray}

Next we consider the ELCS $\overline{\sigma }_{\parallel }(\varphi )$ and $\overline{\sigma }_{\perp }(\varphi )$
for vanishing cyclotron radius, $a_{1},a_{2}\to 0$, i.e. when initially the particles move along the magnetic
field ($v_{01\bot }=v_{02\bot }=0$) which are given by Eqs.~\eqref{eq:van1}-\eqref{eq:fun}. Substitution of
Eq.~\eqref{eq:a48} into Eq.~\eqref{eq:fun} yields
\begin{equation}
\mathcal{F}_{0}\left( \kappa \right) =\frac{2\kappa ^{2}\lambda
^{2}+\varkappa ^{2}+1}{2\left( \varkappa ^{2}-1\right) }\ln \frac{\kappa
^{2}\lambda ^{2}+\varkappa ^{2}}{\kappa ^{2}\lambda ^{2}+1}-1 .
\label{eq:fun1}
\end{equation}%
Thus the ELCS at $a_{1},a_{2}\to 0$ are then given by Eqs.~\eqref{eq:van1} and \eqref{eq:van2}, where the function
$\mathcal{F}_{0}(\kappa )$ is defined by Eq.~\eqref{eq:fun1}.

\begin{figure*}[tbp]
\includegraphics[width=8cm]{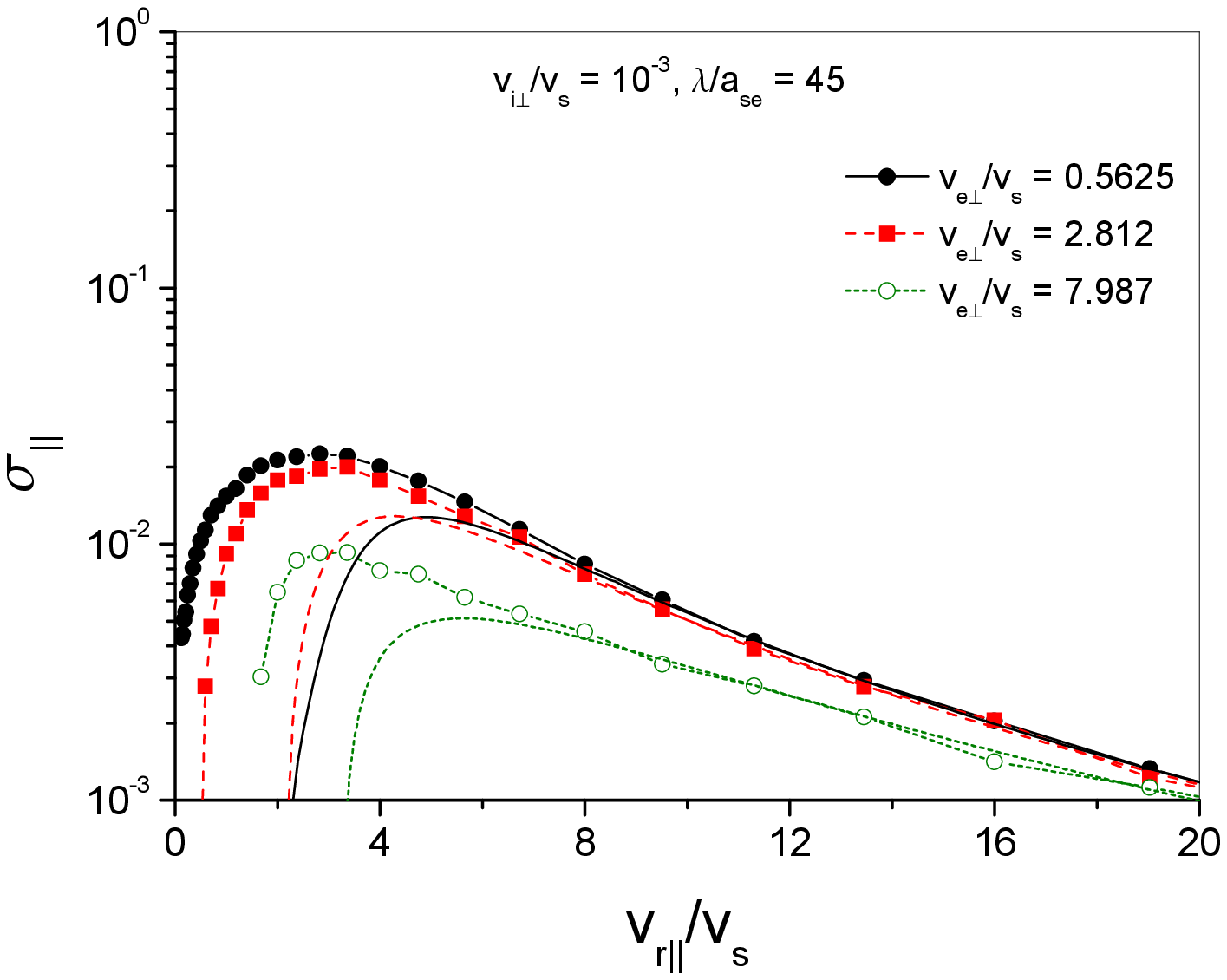} %
\includegraphics[width=8cm]{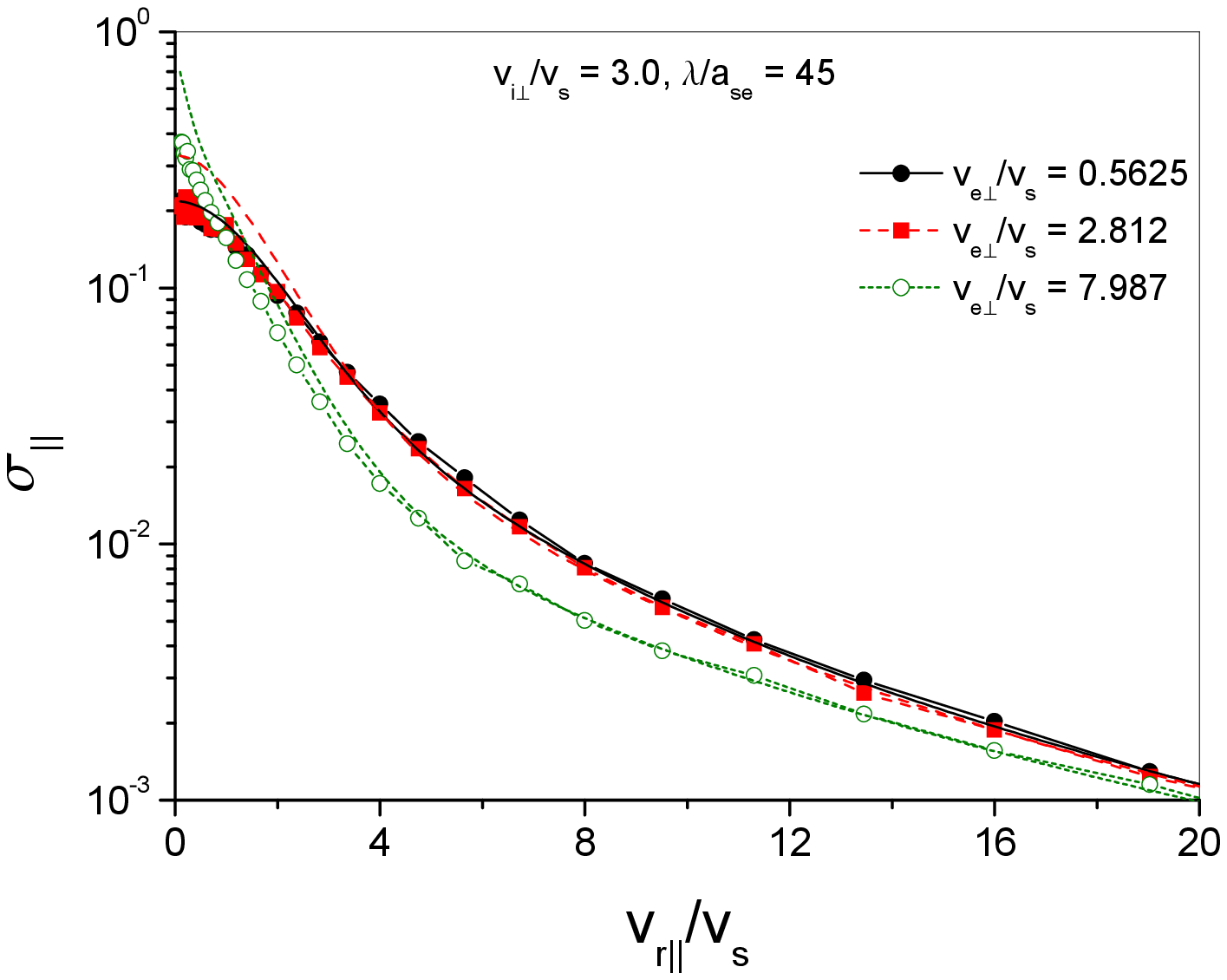}
\includegraphics[width=8cm]{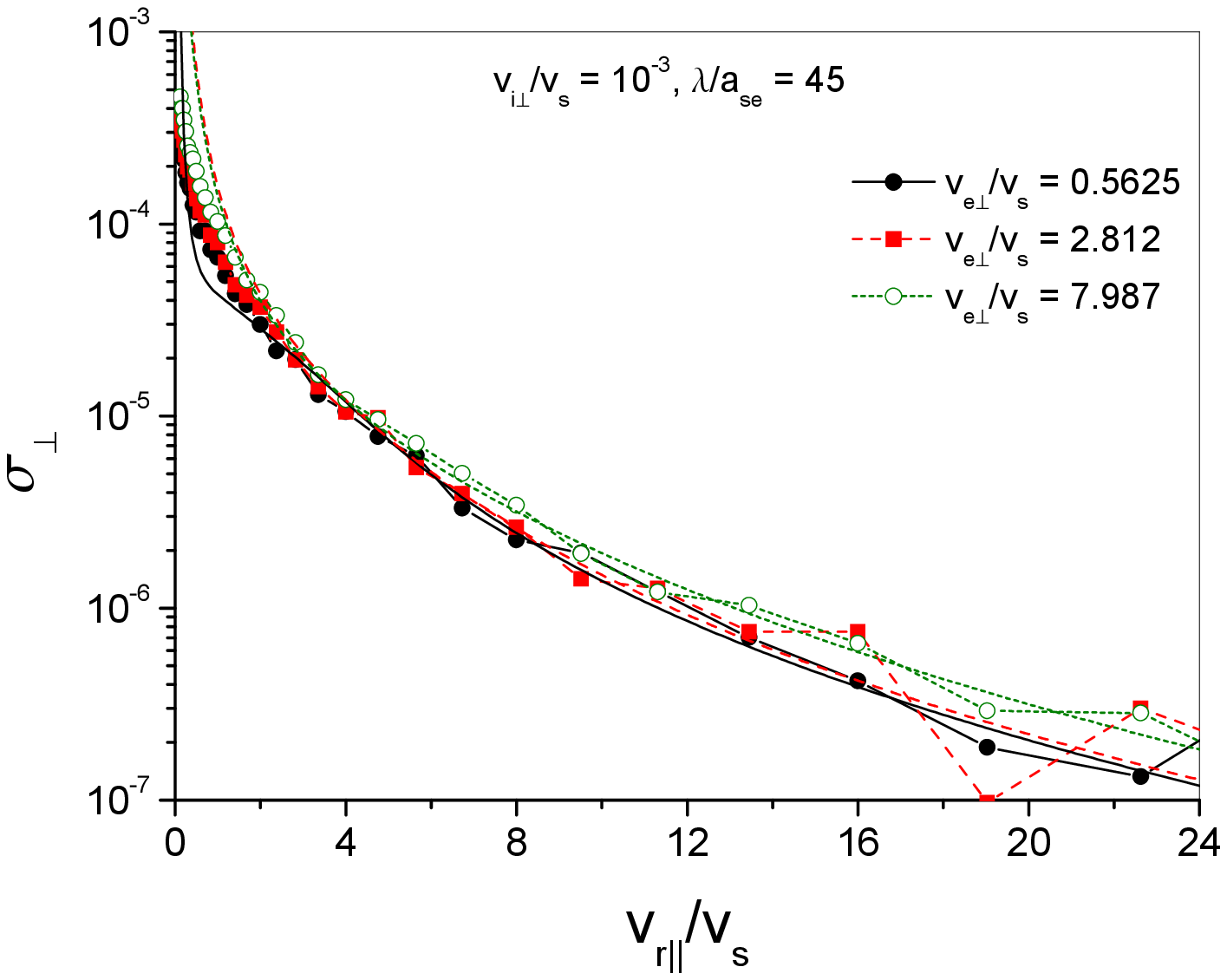}
\includegraphics[width=8cm]{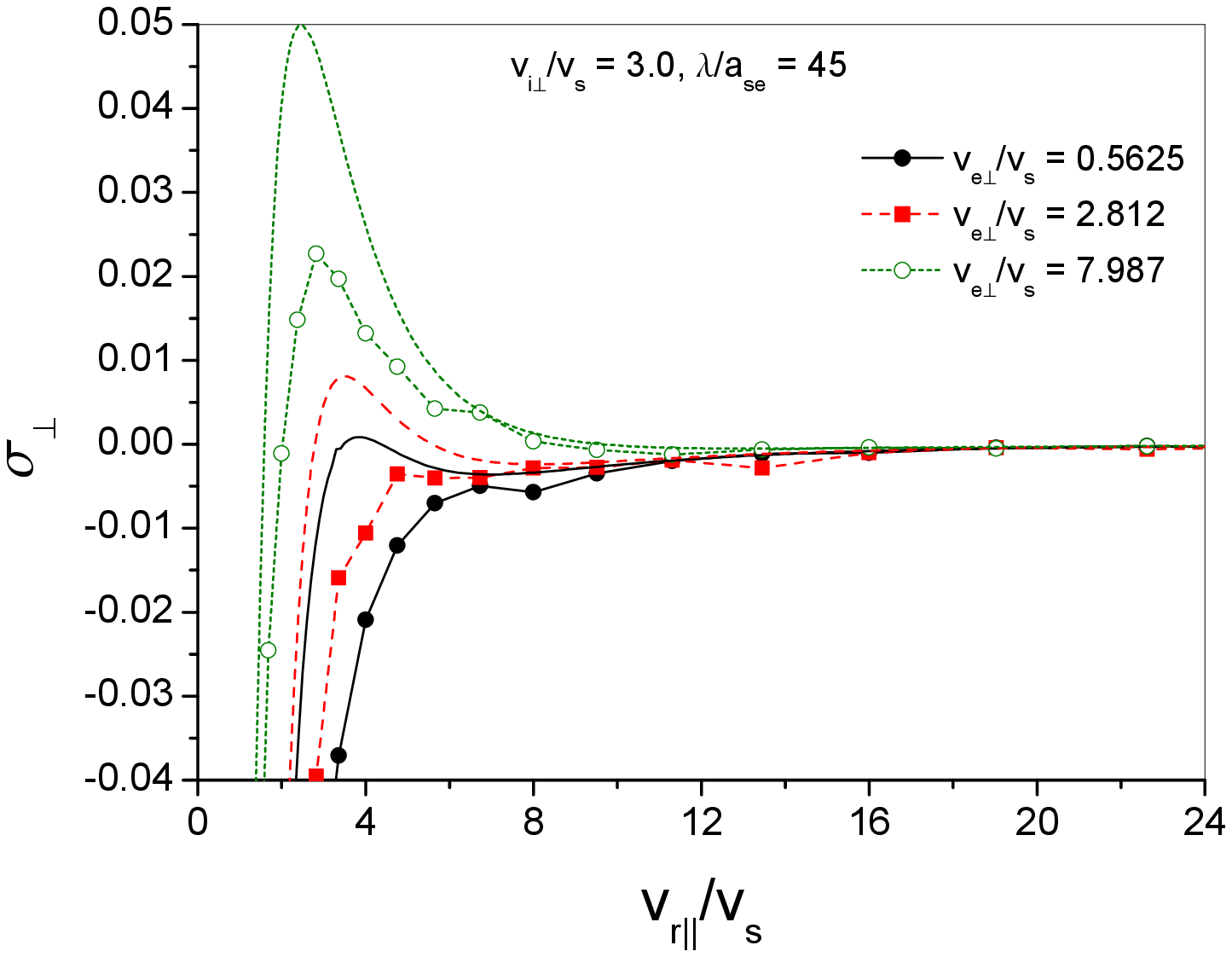}
\caption{(Color online) Top panels, the ELCS $\sigma_{\parallel}$ for electron--ion ($Z=6$ and
$m_2/m_1\simeq 2.2\times 10^{4}$) collision at $v_{i\bot}/v_{s}=10^{-3}$ (left panel) and $v_{i\bot}/v_{s}=3.0$
(right panel). The curves with and without symbols correspond to CTMC simulations and
the second-order perturbative treatment, respectively. Also $\lambda /a_{se}=45$, $v_{e\bot}/v_{s} =0.5625$
(solid lines), $v_{e\bot}/v_{s} =2.812$ (dashed lines), $v_{e\bot}/v_{s} =7.987$ (dotted lines).
Bottom panels, same as in top but for $\sigma_{\bot}$.}
\label{fig:1}
\end{figure*}

Equations~\eqref{eq:van1} and \eqref{eq:van2} with Eq.~\eqref{eq:fun1} are
approximately valid also for finite cyclotron radii $a_{1}$ and $a_{2}$,
assuming that the longitudinal velocity $v_{r\parallel }$ is larger than the
transversal ones, $v_{01\perp }$ and $v_{02\perp }$. Indeed, in this case $%
R(t)\simeq t$ since $\delta _{1},\delta _{2}\gg a_{1},a_{2}$ and in Eqs.~\eqref{eq:a63}
and \eqref{eq:a63x} the transversal ELCS $\overline{\sigma }%
_{\bot }(\varphi )$ can be neglected compared to the longitudinal one. This
indicates that in the high velocity limit with $v_{r\parallel }\gg
v_{01\perp },$ $v_{02\perp }$ the transversal motion of the particles as
well as its cm transversal motion are not important and can be neglected.
Since only the contribution of small $t$ is important the function $R(t)$
is approximated by $R(t) \simeq t$. Using this result
for $R(t)$ from Eq.~\eqref{eq:a63x} and Eqs.~\eqref{eq:van1}
and \eqref{eq:van2} with \eqref{eq:fun1} in the high velocity limit we
obtain within the leading term approximation for the ELCS
\begin{eqnarray}
&&\overline{\sigma }_{\parallel }(\varphi ) \simeq -\frac{2q_{1}^{2}q_{2}^{2}e\!\!\!/^{4}
V_{0\parallel }}{\mu v_{r\parallel }^{3}}\Lambda (\varkappa ) ,  \label{eq:a69b} \\
&&\overline{\sigma }_{\perp }(\varphi ) \simeq -\frac{2q_{1}^{2}q_{2}^{2}e\!\!\!/^{4}}%
{Mv_{r\parallel }^{4}}\left[Q_{1} \Lambda (\varkappa )
+Q_{2}\Lambda _{1}(\varkappa ) \right] ,  \label{eq:a69c}
\end{eqnarray}%
where $Q_{2}=(\lambda ^{2}/2) (\omega _{c1}^{2}-\omega _{c2}^{2})$, $Q_{1}=
(m_{1}/\mu )v_{01\perp }^{2}-(m_{2}/\mu )v_{02\perp }^{2}+((m_2 -m_1)/\mu )v_{01\perp }v_{02\perp }\cos \varphi$,
\begin{eqnarray}
&&\Lambda (u)=\frac{u^{2}+1}{u^{2}-1}\ln u-1 ,  \label{eq:ap5} \\
&&\Lambda _{1}\left( u\right) =1+\frac{1}{u^{2}}-\frac{4}{u^{2}-1}\ln u .
\label{eq:ap6}
\end{eqnarray}%
Note that $\overline{\sigma }_{\parallel }(\varphi )$ is isotropic, i.e., do
not depend on $\varphi $ while $\overline{\sigma }_{\perp }(\varphi )$
contains a term which is proportional to $\cos \varphi $. Also, in the high
velocity limit the parallel ELCS, $\overline{\sigma }_{\parallel }(\varphi )$%
, does not depend on the magnetic field strength. The ELCS decay as $\overline{\sigma }_{\parallel }(\varphi )\sim
v_{r\parallel }^{-3}$ and $\overline{\sigma }_{\bot }(\varphi )\sim
v_{r\parallel }^{-4}$.

Finally we briefly turn to the case of small relative velocity, $v_{r\parallel }\ll v_{01\bot},v_{02\bot}$.
It should be emphasized that Eqs.~\eqref{eq:a63} and \eqref{eq:a63x}, are not adopted
for evaluation of the ELCS at small velocities. For this purpose it is convenient to use an alternative
Bessel-function representation of the ELCS as shown in the Appendix. In addition, it is expected
that the limit of small $v_{r\parallel }$ is the most critical regime for a violation of the perturbation
theory employed here. Therefore explicit analytical expressions in this limit can be useful for an
improvement of the perturbation theory by comparing the analytical results with numerical simulations,
see Sec.~\ref{sec:s4}.

\section{\label{sec:s4}Results. Comparison with simulations}

A fully numerical treatment is required for applications beyond the perturbative regime and for
checking the validity of the perturbative approach outlined above. In the present case of binary
collisions of two arbitrary particles in a magnetic field and with the effective interaction
$U_{\mathrm{R}}(r)$ \eqref{eq:a48} the numerical evaluation of the BC energy loss is very complicated,
but can be successfully investigated by classical trajectory Monte Carlo (CTMC)
simulations.\cite{gzwi99,zwi00,zwi02} In the CTMC method \cite{abr66} the trajectories for the relative
motion between the particles are calculated by a numerical integration of the equations of
motion~\eqref{eq:a1} and \eqref{eq:a2}, starting with initial conditions for the parallel
${v}_{r\parallel }$ and the transverse $\mathbf{v}_{0\perp}=\mathbf{v}_{01\perp}-\mathbf{v}_{02\perp}$
relative velocity. The required accuracy is achieved by using a modified velocity-Verlet algorithm
which has been specifically designed for particle propagation in a (strong) magnetic field,\cite{spr99,zwi08}
and by adapting continuously the actual time-step by monitoring the constant of motion
$E=E_{1}+E_{2}=E_{\mathrm{r}}+E_{\mathrm{cm}}$. The resulting relative deviations of $E$ are of the order
of $10^{-6}-10^{-5}$.

\begin{figure*}[tbp]
\includegraphics[width=8cm]{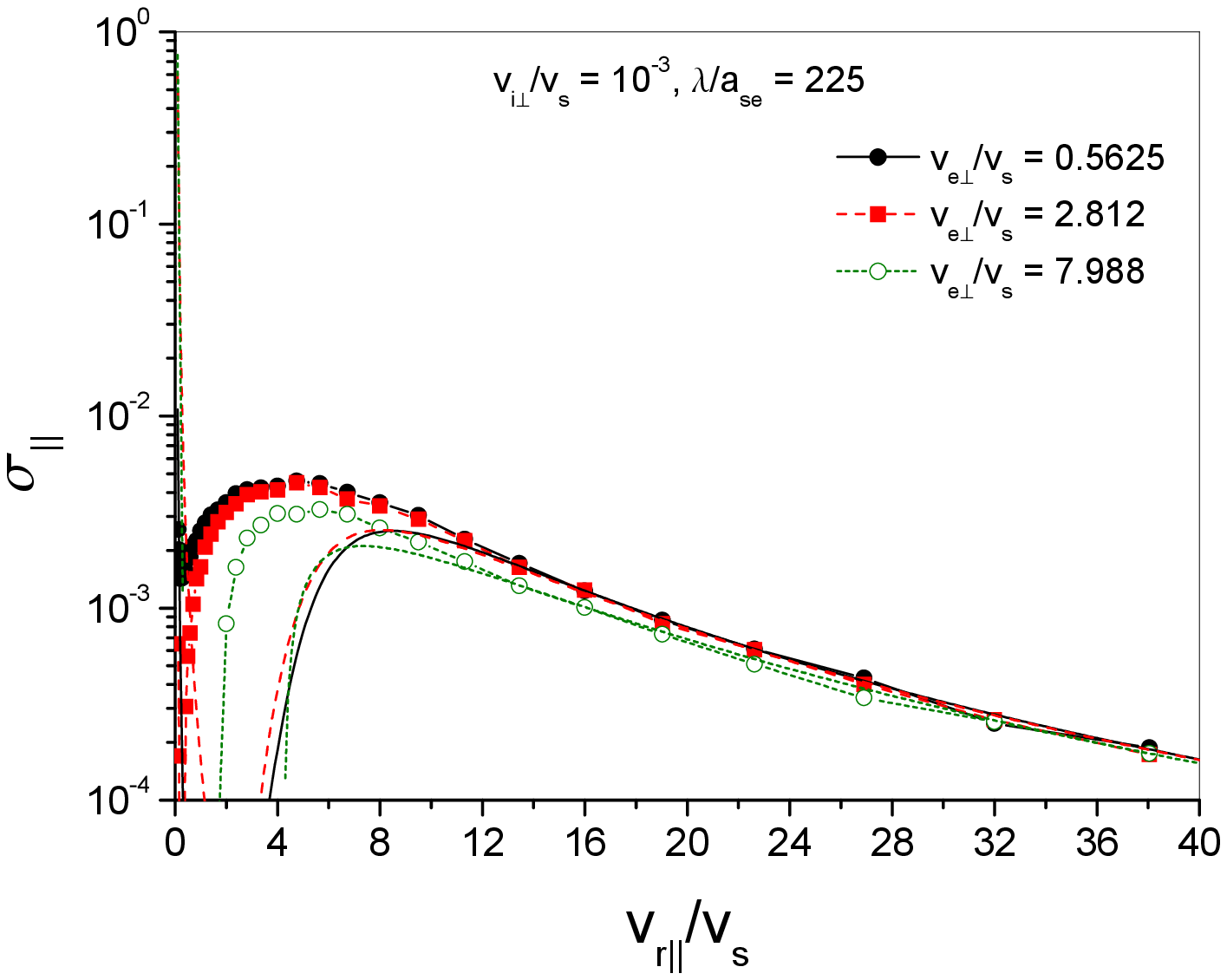}
\includegraphics[width=8cm]{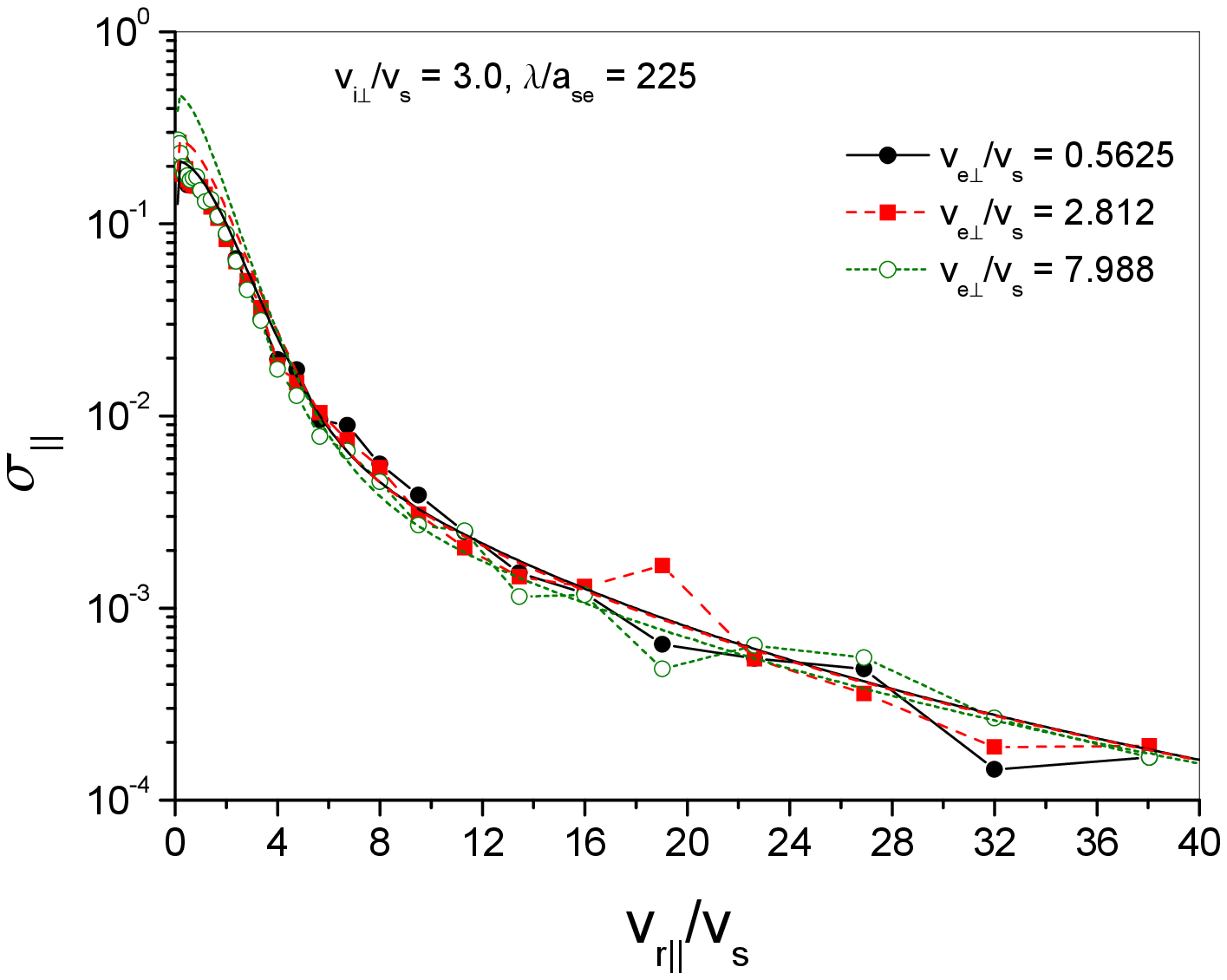}
\includegraphics[width=8cm]{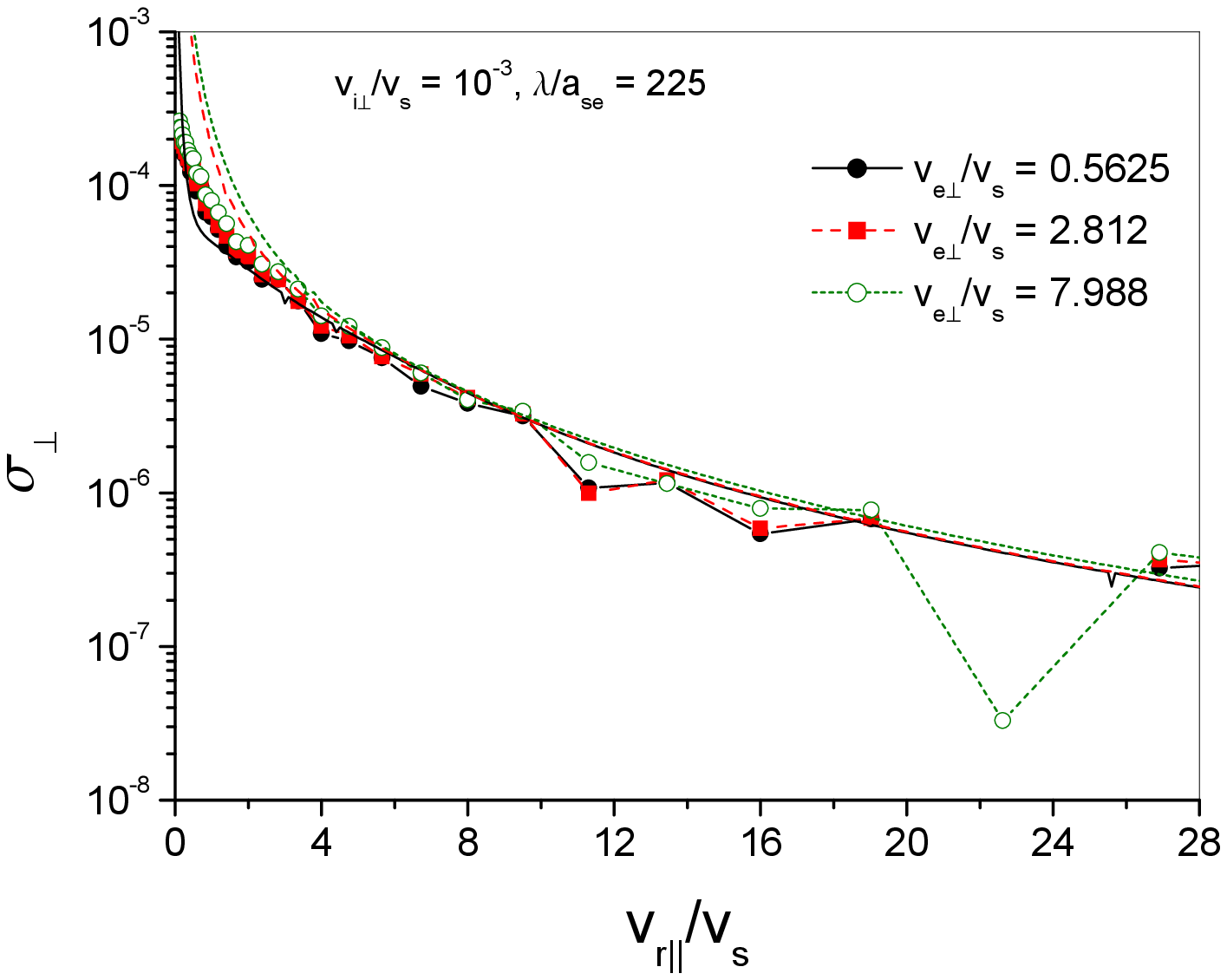}
\includegraphics[width=8cm]{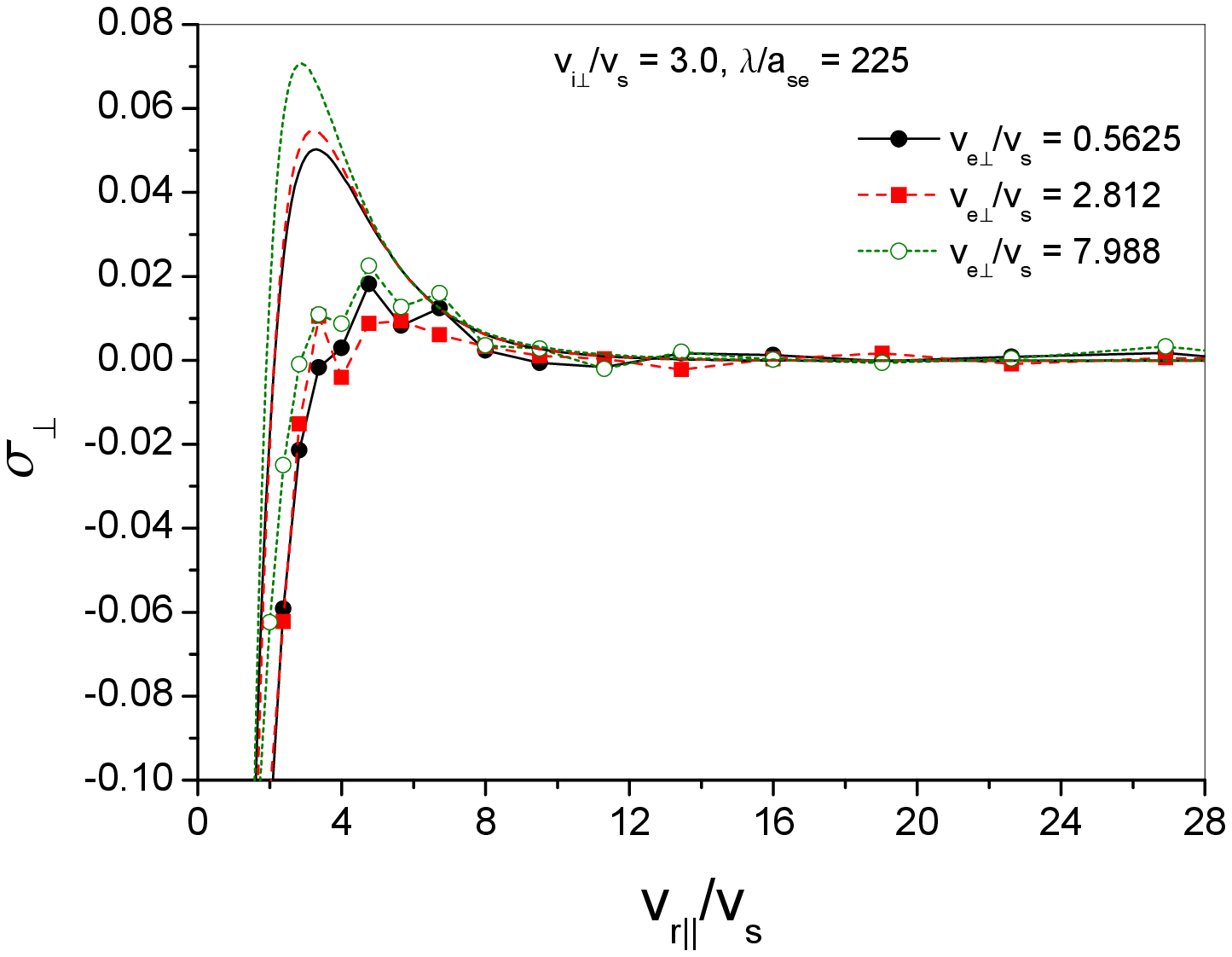}
\caption{(Color online) Same as in Fig.~\ref{fig:1} but for $\lambda /a_{se}=225$ and $v_{e\bot}/v_{s} =7.988$
(dotted lines).}
\label{fig:2}
\end{figure*}

The desired average over the initial phases and the impact parameter $\mathbf{s}$ is performed
by a Monte Carlo sampling \cite{bin97,fis99} of a large number of trajectories with different
initial values. The actual number of computed trajectories (typically $10^{5}-10^{6}$ trajectories)
is adjusted by monitoring the convergence of the averaging procedure. For further details we
refer to Refs.~\onlinecite{ner07,ner09}.

For the forthcoming discussion we put the equation of the relative motion of two particles in a
more appropriate dimensionless form by scaling lengths in units of the screening length $\lambda $
and velocities in units of a characteristic velocity $v_{s}$ defined by $v_{s}^{2}=\vert q_{1}%
q_{2}\vert e\!\!\!/^{2}/\mu \lambda $. This velocity gives a measure for the strength of the Coulomb
interaction with respect to the (initial) kinetic energy of relative motion $\mu v_{r}^{2}/2$. For
$v_{r}<v_{s}$ the kinetic energy is small compared to the characteristic potential energy $\vert %
q_1 q_2 \vert e\!\!\!/^{2}/\lambda $ in a screened Coulomb potential and we expect to be in a
nonperturbative regime. A perturbative treatment on the other hand should be applicable for
$v_{r}\gg v_{s}$.

The scaled version of Eqs.~\eqref{eq:ind1}-\eqref{eq:a2} depends on the four dimensionless parameters
$a_{s1}/\lambda$, $a_{s2}/\lambda$, $m_{2}/m_{1}$ and $\lambdabar/\lambda$, and the initial
conditions with positions scaled in $\lambda$ and velocities in $v_s$. Here $a_{s1} = v_{s}/\omega_{c1}$
and $a_{s2} = v_{s}/\omega_{c2}$ are the cyclotron radii for $v_{1\perp} =v_{2\perp}= v_s$ and the
parameters $a_{s1}/\lambda \propto a_{s2}/\lambda  \propto v_s/B$ represents a measure for the strength
of the magnetic field compared to the strength of the Coulomb interaction [which is $\propto v_s^2$].
The ratio $\lambdabar/\lambda$ describes the amount of softening of the screened interaction at $r\to 0$
with $q_{1}q_{2}e\!\!\!/ ^2 U_{\mathrm{R}}(r\to 0) \to {q_{1}q_{2}e\!\!\!/ ^2}/\lambdabar$.

In the analytical perturbative approach we thus apply the same scaling of length and velocities and
introduce for two particles collisions the dimensionless ELCS
\begin{equation}
\sigma _{\alpha } =-\frac{1}{\mu v_{s}\mathcal{V}_{\alpha}\lambda ^{2}}%
\int_{0}^{2\pi }\frac{d\varphi }{2\pi }\overline{\sigma }_{\alpha }\left(
\varphi \right) ,  \label{eq:sig1}
\end{equation}
where $\alpha =\parallel ,\bot$, $\mathcal{V}_{\parallel}=V_{0\parallel}$ and $\mathcal{V}_{\bot}=v_{s}$.

\begin{figure*}[tbp]
\includegraphics[width=8cm]{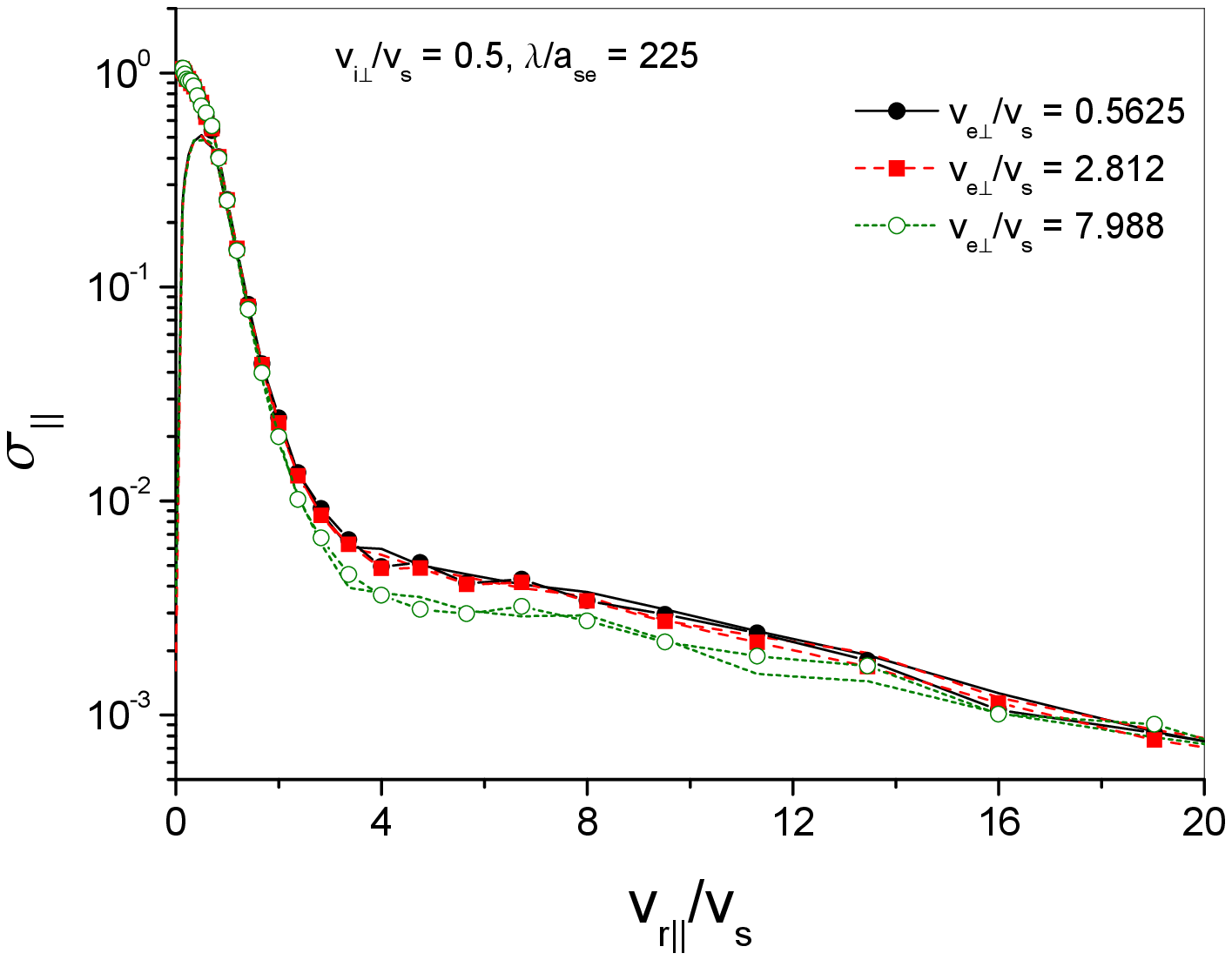} %
\includegraphics[width=8cm]{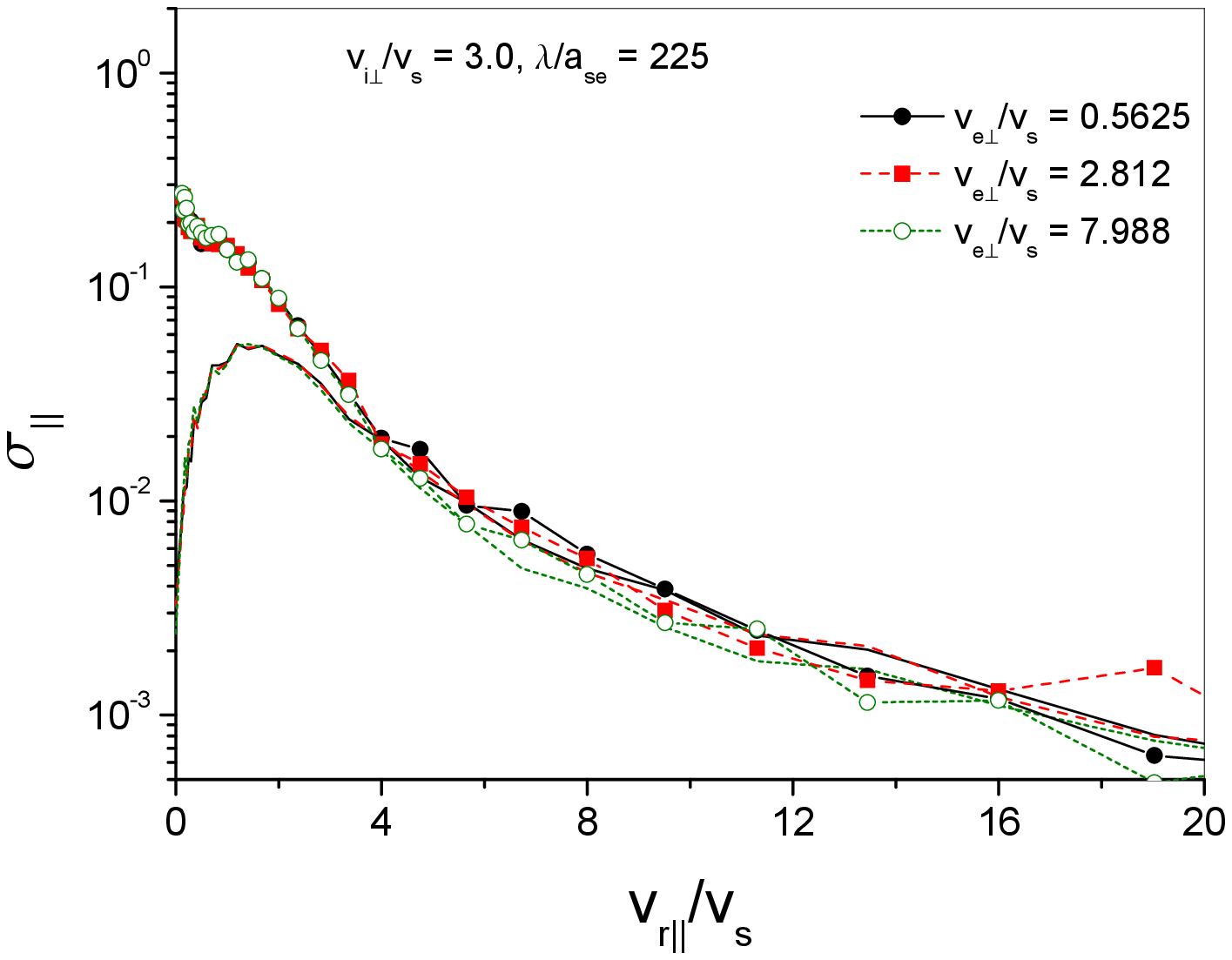}
\includegraphics[width=8cm]{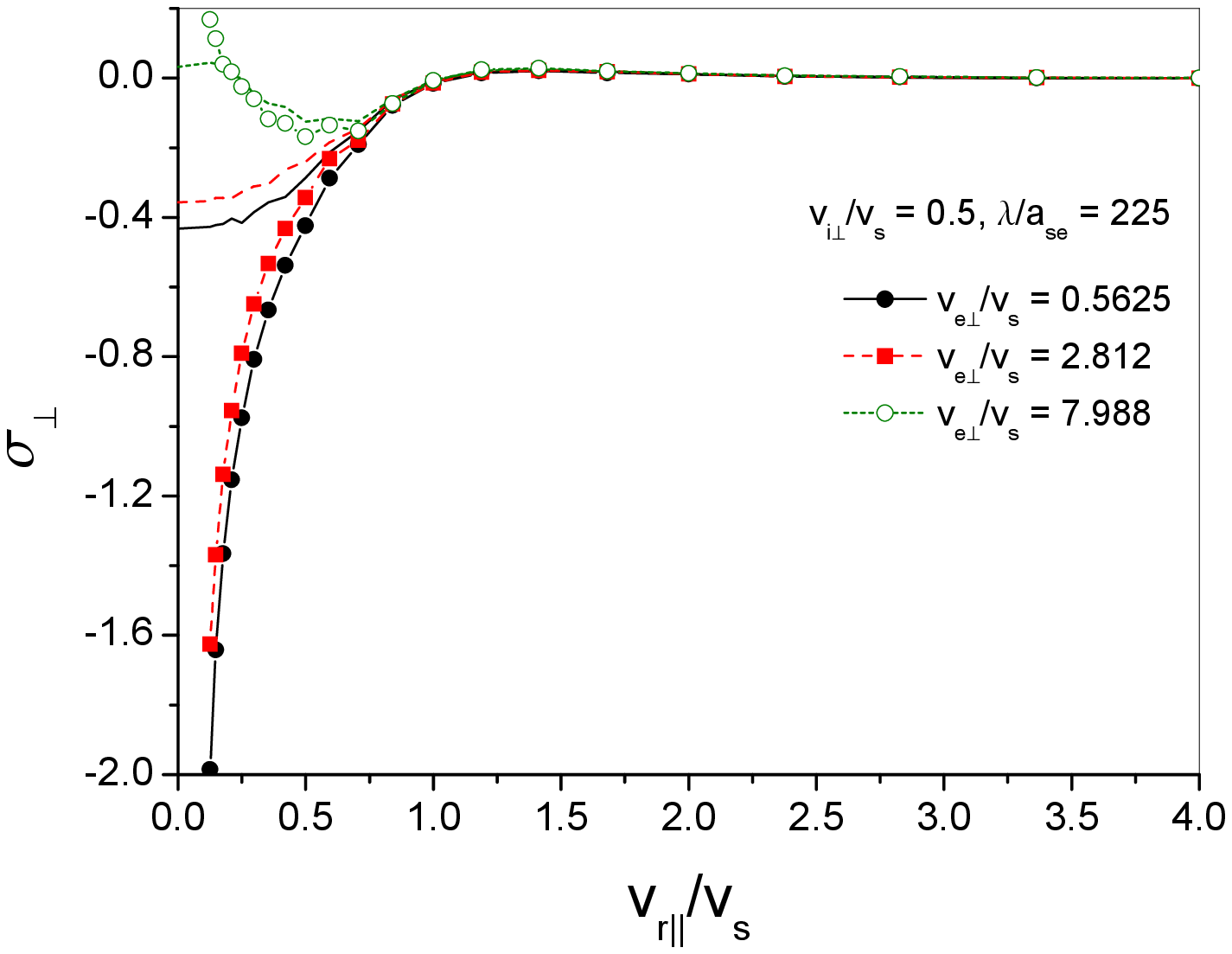}
\includegraphics[width=8cm]{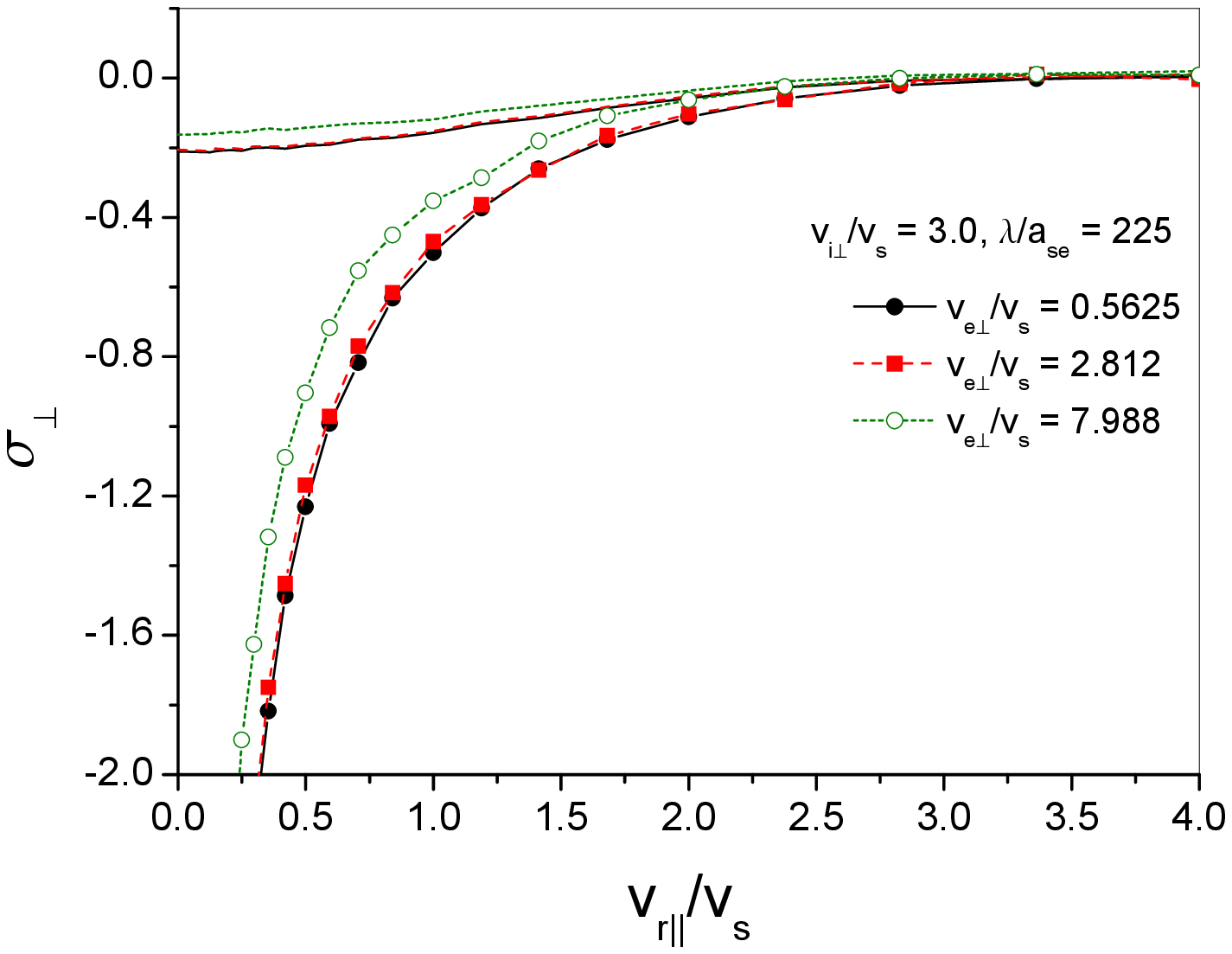}
\caption{(Color online) Top panels, the ELCS $\sigma_{\parallel}$ for electron-ion ($Z=6$ and $m_{2}/m_{1}\simeq 2.2\times 10^{4}$)
collision at $v_{i\bot}/v_{s}=0.5$ (left panel) and $v_{i\bot}/v_{s}=3.0$ (right panel). The curves
with and without symbols correspond to the CTMC simulations with and without ion gyration effects, respectively. Also
$\lambda /a_{se}=225$, $v_{e\bot}/v_{s} =0.5625$ (solid lines), $v_{e\bot}/v_{s} =2.812$ (dashed lines) and $v_{e\bot}/v_{s} =7.988$ (dotted
lines). Bottom panels, same as in top but for $\sigma_{\bot}$.}
\label{fig:3}
\end{figure*}

Next we specify the cutoff parameter $\lambdabar$ which is a measure of softening of the interaction
potential at short distances. As we discussed in previous section the regularization in the potential
\eqref{eq:a48} is sufficient to guarantee the existence of the $s$-integrated energy transfers, but
there remains the problem of treating hard collisions. For a perturbation treatment the change in
relative velocity must be small compared to $v_{r}$ and this condition is increasingly difficult to
fulfill in the regime $v_{r}\to 0$. This suggests a physically reasonable procedure: the potential
must be softened near the origin. In fact the parameter $\lambdabar$ should be related to the de
Broglie wavelength which is inversely proportional to $v_{r}$. Here within classical picture we employ
in a perturbative treatment the dynamical cutoff parameter $\varkappa (v_{r\parallel})=1+\lambda /%
\lambdabar (v_{r\parallel})$,\cite{ner09} where $\lambdabar^{2}(v_{r\parallel})=Cb^{2}_{0}(v_{r\parallel})%
+\lambdabar^{2}_{0}$ with $b_{0}(v_{r\parallel})=\lambda v^{2}_{s}/(v^{2}_{r\parallel}+v^{2}_{0})$,
$v^{2}_{0}=v^{2}_{01\perp}+v^{2}_{02\perp}$.
Here $\lambdabar_{0}$ is some constant cutoff parameter, and $b_{0}(v_{r\parallel})$ is the distance
of closest approach of two charged particles in the absence of a magnetic field. Also in $\lambdabar(v_{r\parallel})$
we have introduced a fitting parameter $C\simeq 0.292$. In Ref.~\onlinecite{ner09} this parameter is
deduced from the comparison of the ELCS \eqref{eq:a43} with an exact asymptotic expression derived in
Ref.~\onlinecite{hah71} for the Yukawa-type (i.e., with $\lambdabar \to 0$) interaction potential. As
has been shown in Ref.~\onlinecite{ner09} the second-order ELCS for electron-electron and electron-ion
(but wihtout gyration of the ion) collisions with dynamical cutoff parameter $\lambdabar(v_{r\parallel})$ excellently agrees
with CTMC simulations at high velocities. The CTMC simulations have been carried out with constant
$\lambdabar =\lambdabar_{0}\ll \lambda$, that is, the interaction is almost Coulomb at short distances.

\begin{figure*}[tbp]
\includegraphics[width=8cm]{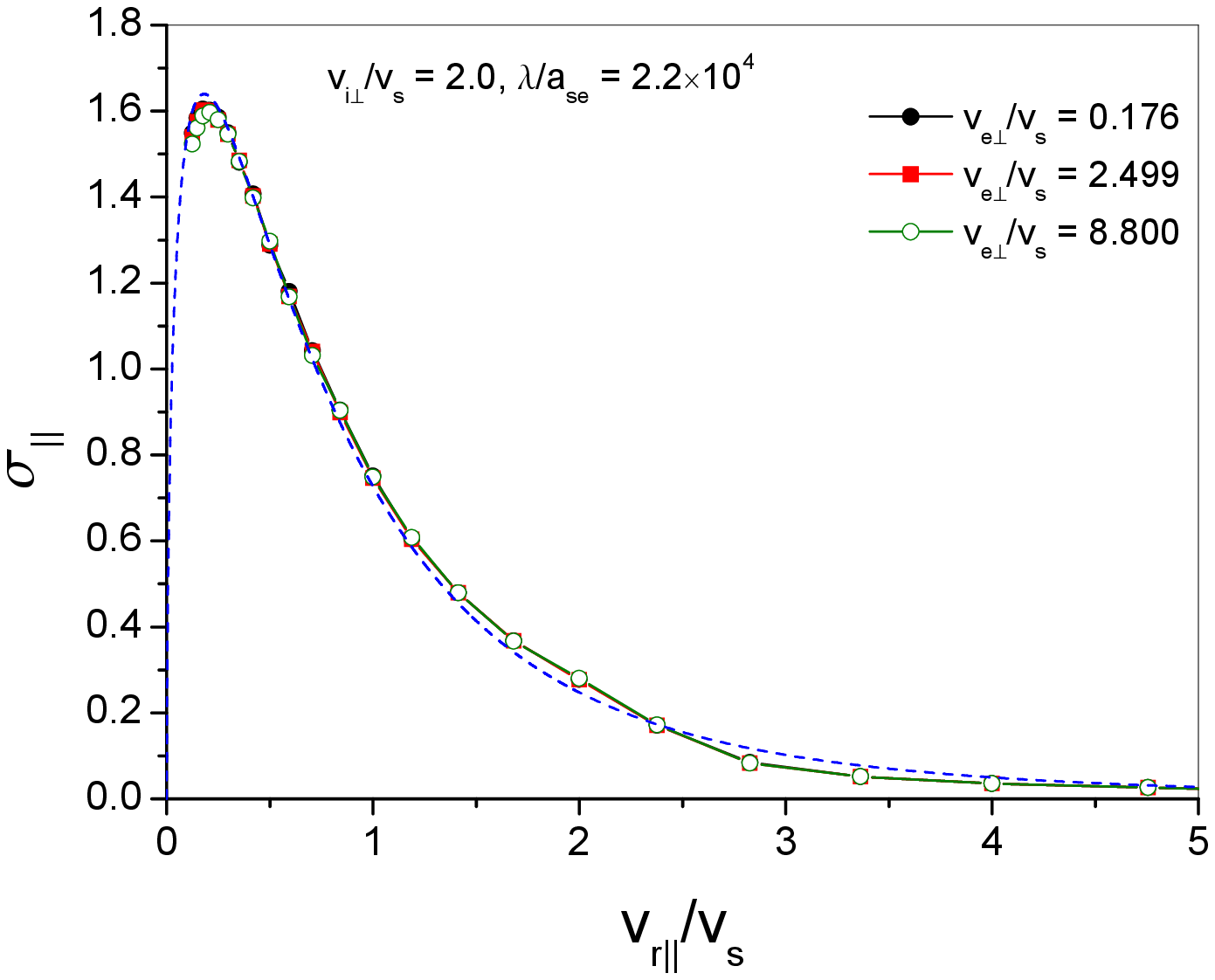}
\includegraphics[width=8cm]{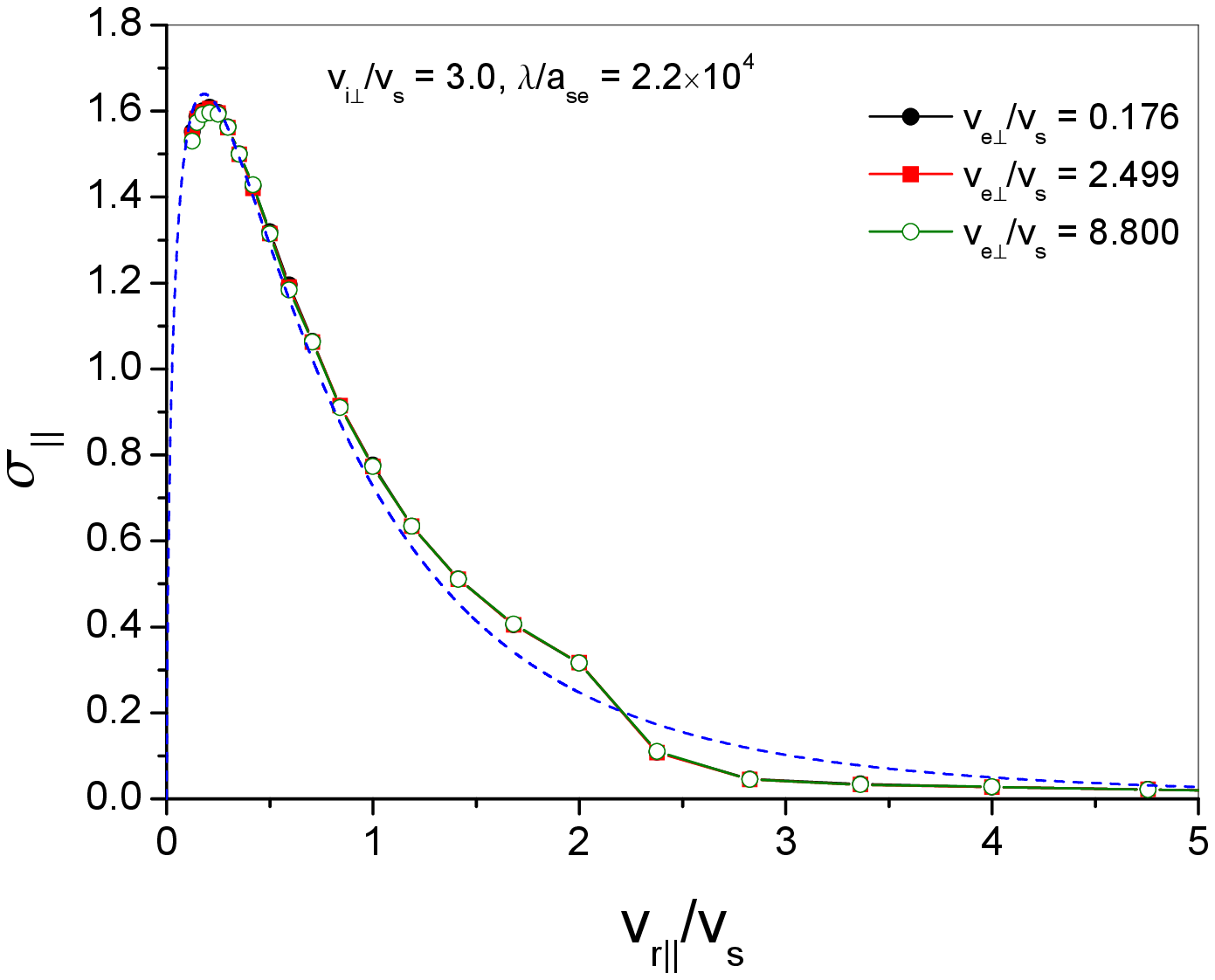}
\caption{(Color online) The ELCS $\sigma_{\parallel}$ for electron--antiproton ($Z=-1$ and
$m_2/m_1 =1836$) collision at $v_{i\bot}/v_{s}=2.0$ (left panel) and $v_{i\bot}/v_{s}=3.0$
(right panel). The curves with symbols correspond to CTMC simulations. Also $\lambda /a_{se}=2.2\times 10^{4}$,
$v_{e\bot}/v_{s} =0.176$
(filled circles), $v_{e\bot}/v_{s} =2.499$ (squares), $v_{e\bot}/v_{s} =8.8$ (open circles).
The dashed curves are obtained employing a model suggested in Ref.~\onlinecite{ner09}.}
\label{fig:4}
\end{figure*}

The ELCS are presented in Figs.~\ref{fig:1}-\ref{fig:4}.
These results are obtained for the BC of an electron (particle 1, $q_{1}=-1$) with an ion (particle 2,
$q_{2}=Z$) in a strong magnetic field. Shown are $\sigma_{\parallel}$ and $\sigma_{\perp}$ as functions of $v_{r\parallel}/v_{s}$
for fixed transverse velocity of the ion $v_{02\bot}=v_{i\bot}$ and the strength of the magnetic field
$\lambda /a_{se}=B/B_{s}$ (with $a_{se}=a_{s1}$, $B_{s}=m_{1}v_{s}/e\lambda$) and varying the transversal
velocity of the electron $v_{01\bot}=v_{e\bot}$. For each triplet of fixed $v_{i\bot}$, $v_{e\bot}$ and
$B$ the cyclotron radii $a_{1}$, $a_{2}$ and $a_{s2}=a_{si}$ can be easily determined. Both the CTMC
and second-order calculations have been done for a regularized potential $U_{\mathrm{R}}$ with
$\nu_{0}=\lambdabar_{0}/\lambda=10^{-4}$. Note that in all cases shown in Figs.~\ref{fig:1}-\ref{fig:4}
we have $a_{se}/\lambda \ll 1$ and the ELCS is not sensitive to this parameter when $a_{se}/\lambda\to 0$
(see, e.g., Fig.~\ref{fig:4}).

Comparisons of the ELCS determined by the CTMC simulations and the second-order perturbative treatment
Eqs.~\eqref{eq:a63}, \eqref{eq:a63x}, and \eqref{eq:sig1} are
presented in Figs.~\ref{fig:1} and \ref{fig:2}. It is clearly observed that in the regimes of large relative velocities the
second order perturbative treatment agrees almost perfectly (i.e.~within the unavoidable numerical fluctuations) with
the CTMC results. In addition in the limit of very large velocities $v_{r\parallel}/v_{s} \gg 1$
the ELCS $\sigma_{\parallel}$ calculated either within perturbation theory or CTMC method with different
strength of the magnetic field and $v_{i\perp},v_{e\perp}$ converge to the same value. This behavior agrees
with the predictions of the asymptotic Eq.~\eqref{eq:a69b} which is independent on $B$ and $v_{i\perp}$,
$v_{e\perp}$. Also the smaller the transversal velocities, the better is the convergence to the regime of
Eq.~\eqref{eq:a69b}. At large relative velocities the CTMC and second-order $\sigma_{\perp}$ shown
in Figs.~\ref{fig:1} and \ref{fig:2} agree with Eq.~\eqref{eq:sig1} with the asymptotic Eq.~\eqref{eq:a69c}. Since
at $v_{r\parallel}/v_{s} \gg 1$ the quantity $\sigma_{\perp}$ behaves as $\sigma_{\perp}\sim v^{2}_{i\perp}$
for fixed magnetic field and $v_{e\perp}$ it will increase with the transverse velocity $v_{i\perp}$ as shown
in Figs.~\ref{fig:1} and \ref{fig:2}. At small velocities with $v_{r\parallel}/v_{s} \lesssim 8$ the second-order treatment
considerably deviates from CTMC simulations. Here the second-order ELCS are given by
approximate expressions \eqref{eq:apb7} and \eqref{eq:apb9} where the parameter $\varkappa$ at small relative
velocities is given by $\varkappa \simeq \varkappa (0)=1+\lambda/\lambdabar (0)$ and $\lambdabar (0)$ is the
dynamical cutoff $\lambdabar (v_{r\parallel})$ at $v_{r\parallel}=0$. Note that at finite cyclotron radii of
the particles the quantity $\lambdabar (0)$ is a constant depending on the value of $\nu_{0}$ and the transversal
velocities. However, for vanishing cyclotron radii (as, e.g. in Eqs.~\eqref{eq:van1} and \eqref{eq:van2}) the
cutoff parameter at small velocities behaves as $\lambdabar/\lambda \sim (v_{s}/v_{r\parallel})^{2}$. The quantity
$(\varkappa -1)^{2}$ involved in Eqs.~\eqref{eq:apb7} and \eqref{eq:apb9} falls as $\sim (v_{r\parallel}/v_{s})^{4}$.
This results in a strong self--cutting at small velocities. Thus employing the cutoff $\lambdabar (v_{r\parallel})$
the second order ELCS $\sigma_{\parallel}$ is strongly reduced and decreases as $\sigma_{\parallel}\sim v^{5}_{r\parallel}$
and $\sigma_{\parallel}\sim v^{3}_{r\parallel}\ln (1/v_{r\parallel})$ at $a_{1}=a_{2}=0$ and $a_{1},a_{2}\neq 0$,
respectively. The second term of $\sigma_{\perp}$ in Eq.~\eqref{eq:apb9} does not contain a term $(\varkappa%
-1)^{2}$ and diverges as $\sigma_{\perp} \sim v^{-2}_{r\parallel}$, see Figs.~\ref{fig:1} and \ref{fig:2}. In this small velocity
regime the second-order perturbative treatment is clearly invalid and a nonperturbative description is required.

To highlight the importance of the ion gyration in the presence of very strong magnetic field we demonstrate
in Fig.~\ref{fig:3} the ELCS obtained with CTMC simulations with and without
ion gyration. That is, in the latter case we assume that $m_{2}\to \infty$ and the ion moves with rectilinear
trajectory with the same velocity component $v_{i\perp}$ as in the case of ion gyration. It is seen that
the ion gyration is important at small $v_{r\parallel}$ and the discrepancy between two approaches increases
with $v_{i\bot}$.

In Fig.~\ref{fig:4} we compared the CTMC results for $\sigma_{\parallel}$ with the model
(dashed lines) given in Ref.~\onlinecite{ner09} for a repulsive ($Z<0$) ion-electron interactions. Here
the ions (antiprotons) are strongly magnetized with $a_{2}/\lambda \simeq 0.17$ (left panel) and $a_{2}/%
\lambda \simeq 0.25$ (right panel). Assuming that the magnetic field is infinitely strong this model
completely ignores the cyclotron motion of the particles and they move along $\mathbf{b}$. It has been
shown\cite{ner09} that for repulsive interaction the magnetic field together with the interaction potential
forms a potential barrier because of the particles motion is effectively one-dimensional. In this case the
relative velocity transfer is $\Delta v_{\parallel }=-2v_{r\parallel }$ which corresponds to a reversion
of the initial motion, i.e. to a backscattering event. Then the energy transfer is $\Delta E_{1}=-2\mu%
V_{0\parallel }v_{r\parallel } \Theta (v_{c}^{2}-v_{r\parallel }^{2})\Theta (s_{m}-s)$, where $\Theta (z)$
is the Heavyside function, $v_{c}^{2}=2\vert q_{1}q_{2}\vert e\!\!\!/^{2}/\mu \lambdabar $ and $s_{m}$ is
determined from equation $|q_{1}q_{2}|e\!\!\!/^{2}U_{R}(s_{m})=\mu v^{2}_{r\parallel}/2$. It is seen that
in Fig.~\ref{fig:4} the agreement with CTMC simulations is quite
satisfactory even for finite cyclotron radii and magnetic field. However, with increasing
$v_{i\bot}$ the CTMC simulations show a more involved picture as, e.g., in Fig.~\ref{fig:4}, right panel,
than the predictions of this simple model. In the CTMC simulations the ELCS shrinks strongly at $v_{r\parallel}%
\gtrsim v_{s}$ with increasing relative velocity and $v_{i\bot}$, $v_{e\bot}$. At strong but finite magnetic
field the hard collisions like backscattering events may also occur but the transverse dynamics of the particles
will reduce the domain of the backscattering events\cite{ner09}. With increasing $v_{r\parallel}$ this domain
will be further shrunk and finally the scattering may occur only in the regime where a strong magnetic field
may strongly reduce the energy transfer.\cite{ner03,ner07,ner09}.

\section{\label{sec:disc}Conclusion}

In this paper we have investigated the binary collisions (BC) of two gyrating charged particles
in the presence of constant magnetic field employing second-order perturbation theory and classical
trajectory Monte Carlo (CTMC) simulations. A case with strongly asymmetric masses of the particles
have been considered in detail. The second-order energy transfers for two--particles collision is
calculated with the help of an improved BC treatment which is valid for any strength of the magnetic
field and involves all cyclotron harmonics of the particles motion. For further applications
(e.g., in cooling of ion beams, transport phenomena in magnetized plasmas) the actual calculations
of the energy transfers have been done with a screened interaction potential which is regularized
at the origin. The use of that potential can be viewed as an alternative to the standard cutoff
procedure.

For checking the validity of the perturbative approach and also for applications beyond the perturbative
regime we have employed numerical CTMC simulations. These CTMC calculations have been performed for a
very strong magnetic field and in a wide range of $v_{r\parallel}$ and for a small regularization parameter,
that is, for an interaction which is rather close to Coulomb at short distances. Within the second-order
treatment we have introduced a dynamic cutoff parameter which substantially improves the agreement of
the theory with CTMC simulations. From a comparison with the nonperturbative CTMC simulations we have found
as a quite general rule which is widely independent of the magnetic field strength that the predictions of
the second-order perturbative treatment are very accurate for $v_{r\parallel}/v_{s}\gtrsim 8$ for all studied
parameters and cases. In contrast, for low relative velocities $v_{r\parallel}/v_{s}\lesssim 8$ the results
obtained from perturbation theory strongly deviate from the CTMC simulations. We have also tested the exact
analytical model derived for repulsive interaction and an infinitely strong magnetic field in Ref.~\onlinecite{ner09}
by comparing it in Fig.~\ref{fig:4} with the CTMC simulations and found that the agreement is rather
satisfactory even for finite (but strong) magnetic fields.

We believe that our theoretical findings will be useful for the interpretation of experimental investigations.
Here, it is of particular interest to study some macroscopic physical quantities on the basis of the presented
theoretical model such as cooling forces in storage rings and traps, stopping power of ion beams as well as
transport coefficients in strongly magnetized plasmas. These studies require an average of the energy or velocity
transfers with respect to the velocity distribution of the electrons. The cooling forces obtained by the perturbative
approach are expected to be quite accurate if the low velocity regime only slightly contributes to the
$\mathbf{v}_{r}$ average over $\langle\Delta E\rangle$. That is, if the typical $v_{r\parallel}$, given by the
maximum of the thermal electron velocity and the ion velocity, are large compared to $v_{s}$, as it is usually
the case for, e.g., electron cooling in storage rings.

\begin{acknowledgments}
H.B.N. is grateful for the support of the Alexander von Humboldt Foundation, Germany. This work was
supported by the Bundesministerium f\"{u}r Bildung und Forschung (BMBF) under contract 06ER9064.
\end{acknowledgments}

\appendix

\section{The energy transfer in a small velocity limit}
\label{sec:ap2}

For the second-order BC treatment the most critical situation is the small
velocity regime where we expect some deviations from the nonperturbative
CTMC simulations. For the improvement of the theoretical approach it is
therefore imperative to investigate the energy transfer in the small
velocity limit, $|v_{r\parallel }|\ll v_{01\bot },v_{02\bot }$, or
alternatively $\delta _{1},\delta _{2}\ll a_{1},a_{2}$. In principle this
limit can be evaluated using the integral representation of the ELCS, Eqs.~\eqref{eq:a63}
and \eqref{eq:a63x}. However, while these expressions are very
convenient to calculate the high velocity limit of the energy transfers (see
Sec.~\ref{sec:s2}) they are not adopted for the evaluation of the small
velocity limit due to the oscillatory nature of the function $R(t)$ at $%
v_{r\parallel }\to 0$. In this Appendix we consider instead an
alternative but equivalent expression for the ELCS. For the axially
symmetric interaction potential the ELCS can be evaluated using Eqs.~\eqref{eq:a42}
and \eqref{eq:a43}. We refer the reader to Refs.~\onlinecite{ner07,ner03,ner09}
for details. The integration of Eq.~\eqref{eq:a42} with respect to the impact
parameter $s$ yields the two-dimensional $\delta $
function, $\delta (\mathbf{k}_{\bot }+\mathbf{k}_{\bot }^{\prime })$, which
combining with $\delta (k_{\parallel }+k_{\parallel }^{\prime })$ in Eq.~\eqref{eq:a42}
yields a three-dimensional $\delta $ function $\delta (\mathbf{%
k}+\mathbf{k}^{\prime })$. The $\mathbf{k}^{\prime }$ integration in the
energy transfer can be then performed exactly. Furthermore it can be shown
that the ELCS is determined by the imaginary part of the integrand in Eq.~(%
\ref{eq:a42}), which is expressed by the functions $\delta \lbrack \zeta
_{n,m}(\mathbf{k})]$ [see, e.g., Eq.~(57) of Ref.~\onlinecite{ner03} for ion-electron
collision]. This allows to perform the $k_{\parallel }$ integration. The
final result reads
\begin{eqnarray}
&&\sigma _{\parallel } =-\frac{2q_{1}^{2}q_{2}^{2}e\!\!\!/^{4}V_{0\parallel }%
}{\mu v_{r\parallel }^{3}}\sum_{n,m=-\infty }^{\infty }\kappa _{nm}\bigg\{
\kappa _{nm}\bigg[ 3\Phi _{nm}(\kappa _{nm})  \nonumber \\
&&+k_{\parallel } \frac{\partial }{\partial k_{\parallel }}\Phi _{nm}(k_{\parallel
}) \bigg]+\frac{\mu n\delta _{1}}{m_{1}a_{1}}\frac{\partial }{\partial a_{1}}%
\Phi _{nm}\left( \kappa _{nm}\right)   \nonumber  \\
&&  +\frac{\mu m\delta _{2}}{m_{2}a_{2}}%
\frac{\partial }{\partial a_{2}}\Phi _{nm}\left( \kappa _{nm}\right)
\bigg\} _{k_{\parallel }=\kappa _{nm}},  \label{eq:apb1}
\end{eqnarray}
\begin{eqnarray}
&&\sigma _{\perp } =\frac{2q_{1}^{2}q_{2}^{2}e\!\!\!/^{4}}{v_{r\parallel
}^{3}}\frac{\vert v_{r\parallel }\vert }{v_{r\parallel }}%
\sum_{n,m=-\infty }^{\infty }\left( \frac{n\omega _{c1}}{m_{2}}-\frac{%
m\omega _{c2}}{m_{1}}\right) \nonumber \\
&&\times \left\{ \kappa _{nm}\left[ 2\Phi _{nm}\left(
\kappa _{nm}\right) +k_{\parallel }\frac{\partial }{\partial k_{\parallel }}%
\Phi _{nm}\left( k_{\parallel }\right) \right] \right.  \label{eq:apb2} \\
&&\left. +\frac{\mu n\delta _{1}}{m_{1}a_{1}}\frac{\partial }{\partial a_{1}}%
\Phi _{nm}(\kappa _{nm}) +\frac{\mu m\delta _{2}}{m_{2}a_{2}}%
\frac{\partial }{\partial a_{2}}\Phi _{nm}(\kappa _{nm})
\right\} _{k_{\parallel }=\kappa _{nm}},  \nonumber
\end{eqnarray}%
where $\kappa _{nm}=\Omega _{nm}/|v_{r\parallel }|=n/\delta _{1}+m/\delta
_{2}$, $\Omega _{nm}=n\omega _{c1}+m\omega _{c2}$, and the function $\Phi
_{nm}(k_{\parallel })$ is defined as
\begin{equation}
\Phi _{nm}(k_{\parallel }) =\frac{(2\pi )^{4}}{4}\int_{0}^{\infty }U^{2}(k_{\parallel },k_{\perp })
J_{n}^{2}(k_{\perp }a_{1}) J_{m}^{2}(k_{\perp}a_{2}) k_{\perp }dk_{\perp } .
\label{eq:apb5}
\end{equation}%
Equations~\eqref{eq:apb1} and \eqref{eq:apb2} are equivalent to the integral representations
\eqref{eq:cross}-\eqref{eq:a43bb} of the ELCS, respectively. Also in Eqs.~\eqref{eq:apb1} and
\eqref{eq:apb2} the function $\Phi_{nm}(k_{\parallel })$ is taken at $k_{\parallel }=\kappa _{nm}$.

For evaluation of the ELCS $\sigma _{\parallel }$ and $\sigma _{\perp }$ at small relative velocities
$v_{r\parallel }$ we note that the terms with $\kappa _{nm}=0$ [in particular the term with $n=m=0$]
do not contribute to the parallel ELCS, $\sigma _{\parallel }$, and they must be excluded from the
summation in Eq.~\eqref{eq:apb1}. Similarly the term with $n=m=0$ must be excluded from the summation
in ELCS $\sigma _{\perp }$. Equation~\eqref{eq:apb2} can be split into two parts with $\kappa _{nm}\neq 0$
and $\kappa_{nm}=0$. Therefore in Eq.~\eqref{eq:apb1} and in the first part of Eq.~\eqref{eq:apb2} (with
$\kappa _{nm}\neq 0$) in the limit of small velocities the quantity $\kappa_{nm}$ becomes very large.
For the regularized interaction potential \eqref{eq:a48} from Eq.~\eqref{eq:apb5} at $k_{\parallel
}a_{1},k_{\parallel }a_{2}\gg 1$ in the leading order we obtain
\begin{equation}
\Phi _{nm}(k_{\parallel }) \simeq \frac{(\varkappa ^{2}-1)^{2}}{\pi ^{2}k_{\parallel }^{8}\lambda ^{6}u_{1}u_{2}}
\left[\ln (\vert k_{\parallel }\vert \lambda )-\frac{17}{12}+\Xi _{nm}\right] . \label{eq:apb6}
\end{equation}%
Here $u_{1}=a_{1}/\lambda $, $u_{2}=a_{2}/\lambda $,
\begin{equation}
\Xi _{nm} =\int_{0}^{\infty }\left[ \pi
^{2}u_{1}u_{2}J_{n}^{2}(u_{1}x) J_{m}^{2}(u_{2}x) -\frac{x^{2}}{(x^{2}+1)^{2}}\right] xdx .
\label{eq:apb6-1}
\end{equation}%
Substituting the expression~\eqref{eq:apb6} into Eqs.~\eqref{eq:apb1} and \eqref{eq:apb2}
in the lowest order with respect to $v_{r\parallel }$ we arrive at
\begin{eqnarray}
&&\sigma _{\parallel }\simeq \frac{10q_{1}^{2}q_{2}^{2}e\!\!\!/^{4}V_{0%
\parallel }v_{r\parallel }^{3}}{\pi ^{2}\lambda ^{6}\mu u_{1}u_{2}}(
\varkappa ^{2}-1)^{2}\sum_{n,m=-\infty }^{\infty }\frac{1}{\Omega
_{nm}^{6}} \nonumber \\
&&\times \left[ \ln \left(\frac{|\Omega _{nm}|\lambda }{|v_{r\parallel }|}%
\right)-\frac{97}{60}+\Xi _{nm} \right] , \label{eq:apb7}
\end{eqnarray}
\begin{eqnarray}
&&\sigma _{\perp } \simeq 4q_{1}^{2}q_{2}^{2}e\!\!\!/^{4}\left\{ \frac{%
3v_{r\parallel }^{4}(\varkappa ^{2}-1)^{2}}{\pi ^{2}\lambda
^{6}u_{1}u_{2}}\sum_{n,m=-\infty }^{\infty }\frac{1}{\Omega _{nm}^{7}} \right. \label{eq:apb9} \\
&&\times \left(\frac{m\omega _{c2}}{m_{1}}-\frac{n\omega _{c1}}{m_{2}}\right) \left[ \ln
\left( \frac{|\Omega _{nm}|\lambda }{|v_{r\parallel }|}\right)-\frac{19}{12}+\Xi _{nm}\right]\nonumber \\
&&\left. +\frac{1}{v_{r\parallel }^{2}}\sum_{k=1}^{\infty }
\left[ \frac{k^{2}p^{2}}{m_{1}a_{1}}\frac{\partial }{\partial a_{1}}\Phi _{kp,kq}\left(
0\right) -\frac{k^{2}q^{2}}{m_{2}a_{2}}\frac{\partial }{\partial a_{2}}\Phi
_{kp,kq}\left( 0\right) \right] \right\} . \nonumber
\end{eqnarray}%
In Eqs.~\eqref{eq:apb7} and \eqref{eq:apb9} the terms with $\Omega _{nm}=0$
in the $n,m$ summations must be excluded. The second term in Eq.~\eqref{eq:apb9}
proportional to $v_{r\parallel }^{-2}$ is the contribution of the
terms with $\kappa _{nm}=0$. Note that this relation requires that $\omega
_{c2}/\omega _{c1}=p/q$ is the rational fraction ($p,q$ are two arbitrary
integers) and $m=-kq$, $n=kp$ with $k=\pm 1,\pm 2,...$ If $\omega
_{c2}/\omega _{c1}$ is not rational fraction the second term in Eq.~\eqref{eq:apb9}
must be omitted. Thus, the small-velocity limit of the ELCS
strongly depends on the parameter $\omega _{c2}/\omega _{c1}=|q_{2}/q_{1}|(m_{1}/m_{2})$.
In addition, Eqs.~\eqref{eq:apb7} and \eqref{eq:apb9} are not valid at $\omega _{c1}=\omega _{c2}$, i.e. in the
case of two symmetrical (e.g., electron-electron) or antisymmetrical (e.g.,
electron-positron) particles collisions. In particular, comparing Eqs.~\eqref{eq:apb7}
and \eqref{eq:apb9} with similar expressions derived for electron-electron
collisions \cite{ner09} we then obtain that at small
relative velocities the ELCS behave here as $\sigma _{\parallel }\sim
v_{r\parallel }^{3}\ln (1/|v_{r\parallel }|)$ and $\sigma _{\perp }\sim
v_{r\parallel }^{4}\ln (1/|v_{r\parallel }|)+C/v_{r\parallel }^{2}$ (where $%
C $ is some constant) while in the electron-electron case $\sigma
_{ee\parallel }\sim v_{r\parallel }^{2}$ and $\sigma _{ee\perp }\sim
v_{r\parallel }^{3}+C_{ee}/v_{r\parallel }^{2}$. This special case with $%
\omega _{c1}=\omega _{c2}=\omega _{c}$ can be evaluated starting from Eqs.~\eqref{eq:apb1}
and \eqref{eq:apb2}. Since $\kappa _{nm}=(n+m)/\delta $
(where $\delta =|v_{r\parallel }|/\omega _{c}$) we introduce a new summation
variable $n\to n-m$. Then the $m$ summation can be performed exactly
using the summation formula for $\sum_{m}m^{\ell}J_{n-m}^{2}(a)J_{m}^{2}(b)$
with $\ell=0,1,2$.\cite{gra80} Similarly one can evaluate the second term of
Eq.~\eqref{eq:apb9} with $p=q=1$.

Finally, we consider the small velocity limits of $\overline{\sigma }%
_{\parallel }(\varphi )$ and $\overline{\sigma }_{\perp }(\varphi )$\ at
vanishing cyclotron radii, $a_{1}=a_{2}=0$ [see Eqs.~\eqref{eq:van1} and
\eqref{eq:van2} with Eq.~\eqref{eq:fun1}],
\begin{eqnarray}
&&\overline{\sigma }_{\parallel }(\varphi ) =\sigma _{\parallel}\simeq -
\frac{q_{1}^{2}q_{2}^{2}e\!\!\!/^{4}V_{0\parallel }v_{r\parallel }}{%
6\lambda ^{4}}(\varkappa ^{2}-1)^{2} \nonumber \\
&&\times \left( \frac{1}{m_{1}\omega _{c1}^{4}}+\frac{1}{m_{2}\omega _{c2}^{4}}\right) ,  \label{eq:apb16} \\
&&\overline{\sigma }_{\perp }(\varphi ) =\sigma _{\perp }\simeq
\frac{q_{1}^{2}q_{2}^{2}e\!\!\!/^{4}v_{r\parallel }^{2}}{6M\lambda ^{4}}%
(\varkappa ^{2}-1)^{2}\left( \frac{1}{\omega _{c1}^{4}}-\frac{1%
}{\omega _{c2}^{4}}\right) .
\label{eq:apb17}
\end{eqnarray}%
In this case $\overline{\sigma }_{\parallel }(\varphi )\sim v_{r\parallel }$
and $\overline{\sigma }_{\perp }\left( \varphi \right) \sim v_{r\parallel
}^{2}$ at small relative velocity $v_{r\parallel }$ [cf. with Eqs.~\eqref{eq:apb7}
and \eqref{eq:apb9}].

\end{document}